\documentclass[lettersize,journal]{IEEEtran}
\usepackage{amsmath,amssymb,amsfonts}
\usepackage{algorithmic}
\usepackage{placeins}
\usepackage{array}
\usepackage{textcomp}
\usepackage{stfloats}
\usepackage{url}
\usepackage{verbatim}
\usepackage{graphicx}
\usepackage{tikz}
\usepackage{adjustbox}
\usepackage{cite}
\usepackage{xcolor}
\usepackage{tabu}
\usepackage{nomencl}
\usepackage{array}
\usepackage{multirow}
\usepackage{subcaption}
\usepackage{tikz}
\usepackage{geometry}
\definecolor{darkblue}{rgb}{9,0,0}
\definecolor{lightyellow}{rgb}{1, 1, 0.8}
\definecolor{lightblue}{rgb}{0.9,0.9,1}
\usepackage{bm}
\usepackage{threeparttable}
\usepackage{tabularx}
\definecolor{lightgreen}{rgb}{0.75, 0.95, 0.75} 
\definecolor{lightgray}{rgb}{0.95, 0.95, 0.95}
\usepackage{soul}

\makenomenclature

\renewcommand{\nompreamble}{The following abbreviations and symbols will be used in forthcoming equations.}
\usepackage{etoolbox}
\renewcommand\nomgroup[1]{%
  \item[\bfseries
  \ifstrequal{#1}{R}{Heavy-Duty Manipulator}{%
  \ifstrequal{#1}{Z}{Control}{%
  \ifstrequal{#1}{E}{EMLA Parameters}{%
  \ifstrequal{#1}{A}{Abbreviations}{}}}}%
]}

\newcommand{\bld}[1]{\mbox{\boldmath $#1$}} 
\usepackage[ruled,vlined,lined,linesnumbered,english]{algorithm2e}

\SetCommentSty{mycommfont}
\SetKwProg{Fn}{Function}{}{}

\begin{document}

\title{System-Level Efficient Performance of EMLA-Driven Heavy-Duty Manipulators via Bilevel Optimization Framework with a Leader--Follower Scenario}

\author{Mohammad Bahari, Alvaro Paz, Mehdi Heydari Shahna, and Jouni Mattila
\thanks{Funding for this research was provided by the Business Finland partnership project ``Future All-Electric Rough Terrain Autonomous Mobile Manipulators'' (Grant No. 2334/31/2022).}
\thanks{All authors are with the Faculty of Engineering and Natural Sciences, Tampere University, 33100 Tampere, Finland.}
	}

\markboth{}%
{Shell \MakeLowercase{\textit{et al.}}: A Sample Article Using IEEEtran.cls for IEEE Journals}


\newpage
\thispagestyle{empty} 
    © 2024 IEEE. Personal use of this material is permitted. Permission from IEEE must be obtained for all other uses, including reprinting/republishing this material for advertising or promotional purposes, collecting new collected works for 
    resale or redistribution to servers or lists, or reuse of any copyrighted component of this work in other works. This 
    work has been submitted to the IEEE for possible publication. Copyright may be transferred without notice, after which this
    version may no longer be accessible.
\newpage 

\maketitle

\begin{abstract}
The global push for sustainability and energy efficiency is driving significant advancements across various industries, including the development of electrified solutions for heavy-duty mobile manipulators (HDMMs). Electromechanical linear actuators (EMLAs), powered by permanent magnet synchronous motors, present an all-electric alternative to traditional internal combustion engine (ICE)-powered hydraulic actuators, offering a promising path toward an eco-friendly future for HDMMs.
However, the limited operational range of electrified HDMMs, closely tied to battery capacity, highlights the need to fully exploit the potential of EMLAs that driving the manipulators.
This goal is contingent upon a deep understanding of the harmonious interplay between EMLA mechanisms and the dynamic behavior of heavy-duty manipulators. To this end, this paper introduces a bilevel multi-objective optimization framework, conceptualizing the EMLA-actuated manipulator of an electrified HDMM as a leader–follower scenario. At the leader level, the optimization algorithm maximizes EMLA efficiency by considering electrical and mechanical constraints, while the follower level optimizes manipulator motion through a trajectory reference generator that adheres to manipulator limits. This optimization approach ensures that the system operates with a synergistic trade-off between the most efficient operating region of the actuation system, achieving a total efficiency of 70.3$\%$, and high manipulator performance. Furthermore, to complement this framework and ensure precise tracking of the generated optimal trajectories, a robust, adaptive, subsystem-based control strategy is developed with accurate control and exponential stability. The proposed methodologies are validated on a three-degrees-of-freedom manipulator, demonstrating significant efficiency improvements while maintaining high-performance operation.
\end{abstract}
\def\abstractname{Note to Practitioners}
\begin{abstract}
The increasing demand for energy efficiency and environmental sustainability is driving the transition in working machine industries toward electric actuation systems. For heavy-duty mobile manipulators (HDMMs), this means replacing old hydraulic and internal combustion engine (ICE)-powered hydraulic actuators with fully-electric electromechanical linear actuators (EMLAs) that run on advanced electric motors, such as permanent magnet synchronous motors. This new actuation mechanism is clean-tech and helps reduce the environmental impact of these working machines. However, one major challenge with EMLA-equipped HDMMs is their limited operation time, constrained by battery capacity. To address this challenge, this paper presents a bilevel optimization framework that focuses on EMLA-driven manipulators, aiming to maximize actuator efficiency while also ensuring that manipulator performance criteria are taken into account. The bilevel optimization aligns the manipulator's trajectory with the EMLAs' most efficient operating regions, improving overall actuation system efficiency to 70.3$\%$. This approach ensures lower energy consumption, which could pave the way to extend operational time for the HDMMs. 
To integrate the control system with the optimization framework for precise trajectory tracking, a robust, adaptive, modular control design is developed in this paper. The modularity of the control system not only reduces the complexity but also simplifies troubleshooting and upgrades, thereby contributing to reduced maintenance costs and facilitating seamless integration into existing industrial setups.
\end{abstract}


\nomenclature[A]{\(\textbf{EMLA}\)}{Electromechanical linear actuator}
\nomenclature[A]{\(\textbf{ID}\)}{Inverse dynamics}
\nomenclature[A]{\(\textbf{MWM}\)}{Mobile working machine}
\nomenclature[A]{\(\textbf{\textit{SO}}(3)\)}{Special orthogonal group in three dimensions}
\nomenclature[A]{\(\textbf{VDC}\)}{Virtual decomposition control}
\nomenclature[A]{\(\textbf{NLP}\)}{Nonlinear programming problem}
\nomenclature[A]{\(\textbf{ICE}\)}{Internal combustion engine}
\nomenclature[A]{\(\textbf{VCP}\)}{Virtual cutting point}
\nomenclature[A]{\(\textbf{HDMM}\)}{Heavy-duty mobile manipulator}
\nomenclature[A]{\(\textbf{HLA}\)}{Hydraulic linear actuator}
\nomenclature[A]{\(\textbf{NE}\)}{Newton--Euler}
\nomenclature[A]{\(\textbf{PMSM}\)}{Permanent magnet synchronous motor}
\nomenclature[A]{\(\textbf{DoF}\)}{Degree-of-freedom}
\nomenclature[A]{\(\textbf{RNEA}\)}{Recursive Newton--Euler algorithm}
\nomenclature[A]{\(\textbf{MCD}\)}{Motor control drive}
\nomenclature[A]{\(\textbf{RDSC}\)}{Robust decomposed system control}
\nomenclature[E]{\(V^r_{ds}\), \(V^r_{qs}\)}{Electric motor d-axis and q-axis voltage in rotor reference frame}
\nomenclature[E]{\(P_{sw}\)}{Motor drive switching loss}
\nomenclature[E]{\(P_{d}\)}{Motor drive conduction loss}
\nomenclature[E]{\(P_{cu}\)}{Stator copper loss of electric motor}
\nomenclature[E]{\(P_{co}\)}{Core loss of electric motor}
\nomenclature[E]{\(P_{hys}\)}{Hysteresis loss in the core of electric motor}
\nomenclature[E]{\(P_{eddy}\)}{Eddy current loss in the core of electric motor}
\nomenclature[E]{\(P_{add}\)}{Additional core loss of electric motor}
\nomenclature[E]{\(P_{mech}\)}{Electric motor mechanical loss}
\nomenclature[E]{\(P_{sc}\)}{Screw mechanism mechanical loss}
\nomenclature[E]{\(P_{EE}\)}{Total power loss during electric to electromagnetic energy conversion}
\nomenclature[E]{\(P_{EM}\)}{Total power loss during electromagnetic to mechanical energy conversion}
\nomenclature[E]{\(\theta_{m}\)}{Mechanical shaft angle of electric motor}
\nomenclature[E]{\(\beta\)}{Inertia coefficient of PMSM}
\nomenclature[E]{\(R_s\)}{Stator resistance of the electric motor}
\nomenclature[E]{\(i^r_{ds}\), \(i^r_{qs}\)}{Electric motor d-axis and q-axis current in rotor reference frame}
\nomenclature[E]{\(L_d\), \(L_q\)}{Electric motor inductance in d-axis and q-axis}
\nomenclature[E]{\(p\)}{Number of pole pairs}
\nomenclature[E]{\(\omega_m\)}{Mechanical angular velocity of electric motor shaft}
\nomenclature[E]{\(J_m\)}{Electric motor standard inertia}
\nomenclature[E]{\(J_c\)}{Electric motor shaft coupling inertia}
\nomenclature[E]{\(J_{GN}\)}{Gearbox inertia}
\nomenclature[E]{\(M_{BS}\)}{Screw mechanism mass}
\nomenclature[E]{\(b_{m}\)}{Viscous coefficient of electric motor shaft}
\nomenclature[E]{\(b_{GB}\)}{friction coefficient between gears}
\nomenclature[E]{\(b_{BS}\)}{Viscous coefficient of screw mechanism}
\nomenclature[E]{\(M_{L}\)}{EMLA mass at the load side}
\nomenclature[E]{\(\Psi_{PM}\)}{Permanent magnet linkage magnetic flux}
\nomenclature[E]{\(\tau_{m}\)}{Electromagnetic torque of electric motor}
\nomenclature[E]{\(f_{x}\)}{Force at the load side of EMLA}
\nomenclature[E]{\(\alpha\)}{Saliency ratio of PMSM}
\nomenclature[E]{\(\mathcal{A}\)}{System matrix of EMLA}
\nomenclature[E]{\(\mathcal{B}\)}{Input matrix of EMLA}
\nomenclature[E]{\(\mathbf{r}\)}{Disturbance vector of system}
\nomenclature[E]{\(\Delta L\)}{Inductance difference in PMSM}
\nomenclature[E]{\(v_{x}\)}{Velocity at the load side of EMLA}
\nomenclature[E]{\(n\)}{Gearbox ratio}
\nomenclature[E]{\(J_{eq}\)}{Equivalent inertia of EMLA at motor side}
\nomenclature[E]{\(b_{eq}\)}{Equivalent friction of EMLA at motor side}
\nomenclature[E]{\(k_{eq}\)}{Equivalent stiffness of EMLA at motor side}
\nomenclature[E]{\(f_{eq}\)}{Equivalent load force ratio of EMLA at motor side}
\nomenclature[E]{\(k_{bearing}\)}{Thrust bearing stiffness}
\nomenclature[E]{\(k_{screw}\)}{Screw mechanism stiffness}
\nomenclature[E]{\(k_{nut}\)}{Ball nut stiffness}
\nomenclature[E]{\(k_{nut}\)}{Thrust tube stiffness}
\nomenclature[E]{\(k_{L}\)}{Linear components stiffness}
\nomenclature[E]{\(k_{\tau 1}\)}{Stiffness between the motor shaft and coupling}
\nomenclature[E]{\(k_{\tau 2}\)}{Stiffness between the coupling and gearbox}
\nomenclature[E]{\(\eta_{\text{EMLA}}\)}{Efficiency of EMLA}
\nomenclature[R]{\(\mathbf{M}_{\mathbf{A}} \)}{Mass matrix}
\nomenclature[R]{\(\mathbf{C}_{\mathbf{A}} \)}{Coriolis and centrifugal matrix}
\nomenclature[R]{\(\mathbf{G}_{\mathbf{A}} \)}{Gravity terms vector}
\nomenclature[R]{\({}^\mathbf{A} \mathbf{R}_{\mathbf{B}} \)}{Rotation matrix between frame \{A\} and \{B\}}
\nomenclature[R]{\({}^\mathbf{A} \mathbf{r}_{\mathbf{AB}} \)}{Skew-symmetric matrix operator from the origin of frame \{A\} to the origin of frame \{B\}}
\nomenclature[R]{\( r_{x} , r_{y} , r_{z} \)}{Distances from the origin of frame \{A\} to the origin of frame \{B\} }
\nomenclature[R]{\( \mathbf{Y}_{\mathbf{A}} \)}{Regressor matrix}
\nomenclature[R]{\( \boldsymbol{\theta}_{\mathbf{A}} \)}{Parameter vector}
\nomenclature[R]{\( \mathbf{{}^{A}}\hat{{\boldsymbol{F}}} \)}{Net force acting on the rigid body with frame \{A\}}
\nomenclature[R]{\({ }^{\mathbf{A}} \mathbf{U}_{\mathbf{B}} \)}{Transformation matrix to convert between frame \{A\} and \{B\}}
\nomenclature[R]{\(\overrightarrow{\mathbf{f}_{\mathbf{A}}}\)}{Force vector applied to a rigid body with attached frame \{A\}}
\nomenclature[R]{\(\overrightarrow{\mathbf{m}_{\mathbf{A}}}\)}{Moment vector applied to a rigid body with attached frame \{A\}}
\nomenclature[R]{\(\overrightarrow{\mathbf{v}_{\mathbf{A}}}\)}{Linear velocity vector applied to a rigid body with attached frame \{A\}}
\nomenclature[R]{\(\overrightarrow{\mathbf{\omega}_{\mathbf{A}}}\)}{Angular velocity vector applied to a rigid body with attached frame \{A\}}
\nomenclature[R]{\(\mathcal{T}\)}{Discretized time}
\nomenclature[R]{\(t_0\)}{Initial time}
\nomenclature[R]{\(t_M\)}{Final time}
\nomenclature[R]{\(t_0\)}{Initial time}
\nomenclature[R]{\(\boldsymbol{w}\)}{Weight vector of optimization problem}
\nomenclature[R]{\(\boldsymbol{w}\)}{Weight vector of optimization problem}
\nomenclature[R]{\(\boldsymbol{q}, \boldsymbol{\dot{q}}, \boldsymbol{\ddot{q}}\)}{Configuration vector and its time derivatives}
\nomenclature[R]{\(\boldsymbol{B}(t)\)}{Basis functions of the B-spline}
\nomenclature[R]{\(\boldsymbol{c}\)}{Control points}
\nomenclature[R]{\(\boldsymbol{q}_L, \boldsymbol{q}_U\)}{Lower and upper limits of the joints position}
\nomenclature[R]{\(\boldsymbol{\dot{q}}_L, \boldsymbol{\dot{q}}_U\)}{Lower and upper limits of the joints velocity}
\nomenclature[R]{\(\boldsymbol{f}_{xL}, \boldsymbol{f}_{xU}\)}{Lower and upper limits of the joints force}
\nomenclature[R]{\(\boldsymbol{v}_{xL}, \boldsymbol{v}_{xU}\)}{Lower and upper limits of the joints velocity}
\nomenclature[R]{\(\boldsymbol{q}_I, \boldsymbol{q}_F\)}{Initial and final joints position}
\nomenclature[R]{\(\boldsymbol{\dot{q}}_I, \boldsymbol{\dot{q}}_F\)}{Initial and final joints velocity}
\nomenclature[R]{\(\boldsymbol{\psi}(t,\boldsymbol{c})\)}{Optimization criteria}
\nomenclature[R]{\(f(.)
\)}{Cost function of inner-level optimization}
\nomenclature[R]{\(\boldsymbol{g}(.)
\)}{Constraints of inner-level optimization}
\nomenclature[R]{\(F(.)
\)}{Cost function of outer-level optimization}
\nomenclature[R]{\(\boldsymbol{G}(.)
\)}{Constraints of outer-level optimization}
\nomenclature[R]{\(\bld{\xi}^{*}\)}{Optimal argument}
\nomenclature[R]{\(n_a\)}{Manipulator number of degrees-of-freedom}

\printnomenclature

\section{Introduction}
\subsection{Background and Context}
\label{background}
\IEEEPARstart{T}{he} harmful effects of greenhouse gas emissions on our ecosystems are clear, and international agreements, such as the 2015 Paris Agreement \cite{delbeke2019paris}, emphasize the urgency to reduce carbon dioxide (CO$_2$) emissions. These agreements highlight the importance of transitioning to electricity as the primary energy source in different industries \cite{borlaug2021heavy}. The shift toward decarbonization and utilizing renewable energies has also influenced the machinery industry, with a growing focus on manufacturing zero-emission mobile working machines (MWM) \cite{fassbender2024energy}.
Contemporary industrial operations are heavily reliant on a diverse range of working machines. Consequently, the global construction and mining machinery industry has witnessed sustained growth in response to expanding urban development. Heavy-duty mobile manipulators (HDMMs) are categorized as MWMs, where the manipulator, characterized as a serial-link device, is mounted on the mobile platform. The primary function of these HDMMs is to assist operators in tasks ranging from activities such as heavy lifting and demolition to more intricate processes such as precision assembly and the delicate handling and placement of objects. 
At the moment, these HDMMs mostly rely on hydraulic linear actuators (HLAs) powered by an internal combustion engine (ICE), a setup that conflicts with modern environmental standards and sustainability goals, to convey power and to generate linear motion. 
While HLAs powered by a battery and electric motor to drive the pump represent a notable stride toward electrification \cite{sakama2022characteristics}, they still suffer from suboptimal energy efficiency due to the additional energy conversion stage from electric power to hydraulic flow and then to mechanical motion \cite{yan2023energy,manring2013efficiency,chipka2015efficiency,pustavrh2023comparison}.
By offering several benefits over HLAs, electromechanical linear actuators (EMLAs) are poised to play a transformative role in the advancement of fully electrified and intelligent HDMMs, marking a critical juncture in this ongoing shift \cite{pustavrh2023comparison,tanaka2013comparative,katsushi_furutani_2022,bolam2018review,qiao2018review}. Advancements in motor technology, particularly the widespread adoption of permanent magnet synchronous motors (PMSMs) \cite{8063879,10016655,10476344}, have offered a remarkable improvement in torque density, low cogging torque, and allowing EMLAs to pack more power into a smaller footprint \cite{9269466,tootoonchian2016cogging}. This, coupled with their outstanding efficiency, wide speed range, and accurate controllability, translates to increased productivity and operational time \cite{lu2021energy,986441,10305210,diesen_fredrik_2021,yamasaki2019electric}. Additionally, they present an opportunity for energy recovery during the duty cycles of lowering loads and restoring energy back to the source \cite{Qi2020energyrecovery,mazzoleni2021electro}, offering a compelling emission-free and efficient solution.

Several studies have covered the evolving landscape of electrification and EMLA optimization topics. For example, in \cite{bissal2015modeling}, a multi-field coupling model encompassing mechanical, electromagnetic, and thermal aspects for an ultra-high-speed electromechanical actuation system is developed. Also \cite{bissal2015modeling} carries out a sensitivity analysis on various parameters, including driver shape, motor coil size, and shell material, leading to a system optimization design. Similarly, \cite{persson2015framework} optimizes three different actuation methods—mechanical, pneumatic, and hydraulic—for the balancing mechanism of industrial robots, focusing on dynamic characteristics, weight, and volume. However, the model does not explore the interrelationship between parameters. Another paper \cite{zheng2023investigations} designs an integrated electromechanical actuator module with a focus on reducing weight, optimizing performance, and addressing heat dissipation through thermal modeling. It also develops models for parameter estimation and multi-objective optimization. A further paper \cite{badrinarayanan2018electro} discusses EMLAs with roller screws, which outperform traditional ball screws in high dynamic load applications. It focuses on optimizing actuator performance and motor sizing, and incorporates safety features such as an electromagnetic clutch for enhanced reliability.

\subsection{Research Gap and Motivations of the Study}
\label{Motivations}
While existing research has made notable progress in optimizing various facets of electric actuation systems, these efforts have primarily focused on isolated aspects within single-level optimization frameworks. These approaches, though valuable, often overlook the crucial interaction between the actuator and the complex dynamics of the systems they drive, which play a major role particularly in heavy-duty applications. As a result, there remains a significant gap in fully optimizing the balance between energy efficiency and operational performance in electrified HDMMs.
Addressing this gap requires a comprehensive understanding of electric actuator parameters, a deep grasp of the targeted mechanisms, and precise control of key variables. Given the limited energy resources in electrified HDMMs, primarily dependent on battery capacity, prioritizing efficient energy usage is essential to extend operating range. Achieving optimal operation of EMLA-driven manipulator necessitates a thorough understanding of energy conversions within EMLAs, the dynamic and kinematic properties of the manipulator, and the integration of motion-generation algorithms and control decision-making \cite{6570761}. The principal challenge in system-level optimization of electrified HDMMs lies in the intricate interplay between the performance of the manipulator and the efficient operation of the EMLA, with the objective of realizing a harmonious and optimal synergic operation. This interaction significantly impacts both energy efficiency and operational performance, yet single-level optimization strategies often fall short by focusing on either the manipulator's trajectory or the actuator's efficiency in isolation. Moreover, even multi-objective optimization struggles to account for the hierarchical and nested decision-making processes inherent in these systems. This narrow focus overlooks the crucial interdependencies, leading to suboptimal outcomes. 
\subsection{Contributions, Methodology, and Structure of the Paper}
This paper addresses the challenges outlined in Section \ref{Motivations} by presenting a bilevel multi-objective optimization through a leader--follower scenario. At the leader level, the optimization algorithm maximizes the actuation system efficiency by taking both electrical and mechanical constraints of EMLAs into account. Concurrently, the follower level refines manipulator motion through a trajectory reference generator that optimizes the performance criteria of the manipulator with respect to its dynamic and kinematic limits. This structured optimization methodology achieves a harmonious trade-off between operating within the most efficient regions of the actuation system and maximizing the manipulator's performance criteria through a coordinated decision-making process. In order to follow the generated optimal trajectory, this paper proposes a robust adaptive modular control strategy. This control system guarantees precise trajectory tracking with exponential stability. 

The paper begins with the evaluation of electrified actuator performance in Section \ref{section:energy_conversion} and investigation of critical energy conversion stages within an EMLA in \eqref{equation:dq0voltage}-\eqref{torque_force}, as well as the power losses in different components, as described in (6)-(12). In Section \ref{section:state_space_EMLA}, the governing dynamic of an EMLA is explained in \eqref{angularvelocity}-\eqref{linear_components_stiffness} and the state space model of the investigated mechanism is obtained and linearized through \eqref{state_space_2}-\eqref{matrix_R} to obtain the efficiency maps of utilized electric actuators, using \eqref{system_efficiency} that is presented in Section \ref{section:energy_conversion}. The paper continues with kinematic and dynamic analysis of the studied heavy-duty manipulator in Section \ref{kinematics_and_dynamic}. The virtual decomposition control (VDC) approach is employed, whereby its fundamental concepts are described in Section \ref{fundamental_vdc} and the mathematical modeling of the manipulator is developed in Section \ref{mathematical_modeling_HDMM}.
To generate optimal trajectories for the manipulator, this paper adopts a trajectory-optimization framework based on direct collocation with B-spline curves in Section \ref{section:trajectory_optimization}. This is achieved using a recursive Newton--Euler algorithm (RNEA) specifically developed for closed kinematic chains through \eqref{time}-\eqref{equation:rnea} and a finite-dimensional nonlinear programming problem (NLP) to form the optimization level with targeting the manipulator through \eqref{equation:costf2}-\eqref{equation:criteria}. 
In Section \ref{section:bilevel_optimization}, a bilevel multi-objective optimization method is employed to formulate the system-level problem as a leader–follower scenario, where the manipulator trajectory generator optimizes decisions according to the criteria defined in \eqref{equation:minimumeffort}-\eqref{equation:minimumpower} of Section \ref{section:trajectory_optimization}, which are the effort and power delivered by the joint pistons in this study, while also considering the decision made by the EMLA criterion that aims to maximize the energy conversion ratio from the battery to the piston and is defined as \eqref{system_efficiency} in Section \ref{section:energy_conversion}. The proposed approach not only promises to generate optimal trajectories aligned with the manipulator’s objectives but also ensures that the EMLAs installed in each joint operate at their highest possible efficiency at the same time. A merit of this approach lies in its capability to resolve the scaling problem in multi-objective optimization. This is accomplished by identifying optimal weights that ensure the maximum efficiency of EMLAs as per \eqref{eq:costf2__}-\eqref{outer_obj}. Subsequently, by developing the control strategy proposed in \cite{heydari2024robust}, which addresses a one-degree-of-freedom (1-DoF) mechanism, a generic robust, adaptive, subsystem-based control is developed for an $n_a$-DoF robotic manipulator in Section \ref{section:control} to achieve precise tracking of optimal trajectories in the actuator, attaining exponential stability in the presence of the external effects on the manipulator and EMLA model, such as manufacturing variations, environmental factors, and sensor inaccuracy. This control approach decomposes manipulator systems into subsystems to control the linear motions of EMLAs and the currents of motors through \eqref{equation:39}-\eqref{equation: 45}. The stability analysis of the proposed control system is proved through \eqref{equation: 450}-\eqref{5400}, in Section \ref{sec:stability}. Finally, in Section \ref{case_study_control}, the control method is applied for the studied 3-DoF manipulator to show the effectiveness of the trajectory tracking.

\begin{figure*} [tbp]
	\centering
	\includegraphics[trim={0.0cm 0.0cm 0.0cm 0.0cm},clip,width=\linewidth]{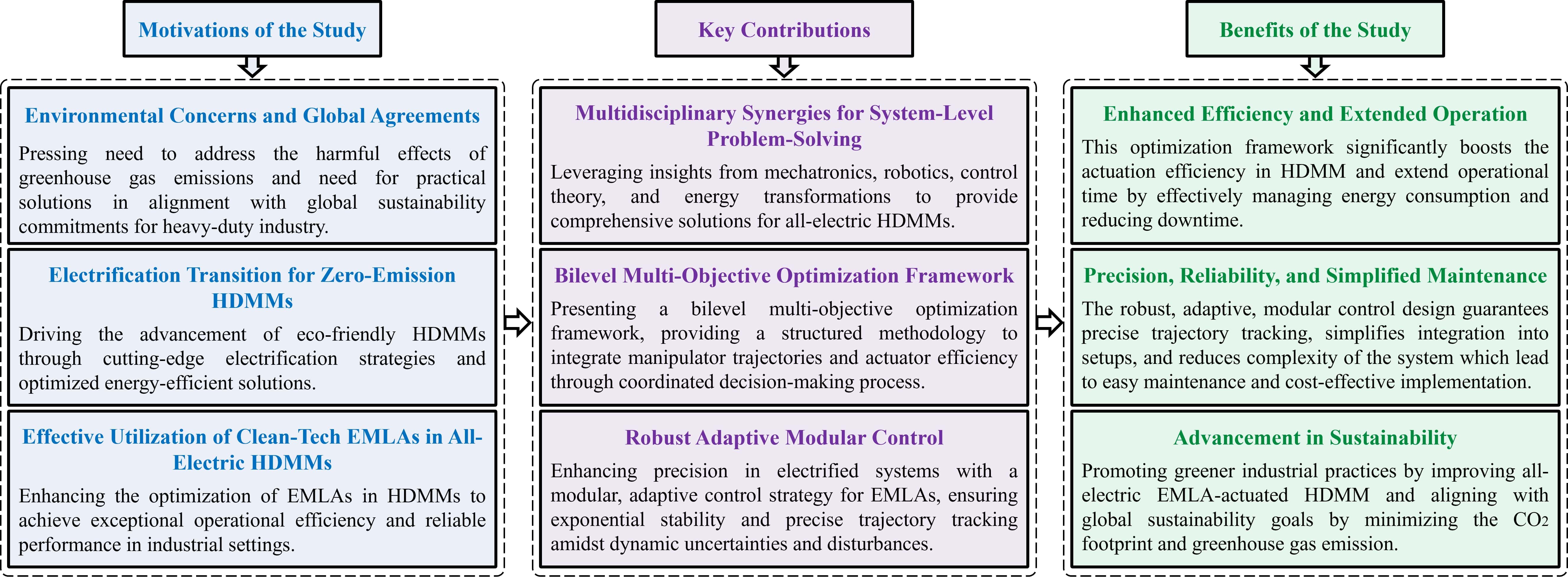}
	\caption{Motivations, key contributions, and benefits of the study}
	\label{Motivations_contributions_benefits}
\end{figure*}

Fig. \ref{Motivations_contributions_benefits} provides a summary of the study’s motivations, the significant contributions, and the benefits of the approach used in this paper to enhance the performance of the all-electric HDMMs.
\FloatBarrier
\section{Energy Conversions and Losses of EMLAs}
\label{section:energy_conversion}
\begin{figure} [tbp]
	\centering
	\includegraphics[trim={0.0cm 0.0cm 0.0cm 0.0cm},clip,width=8.5cm]{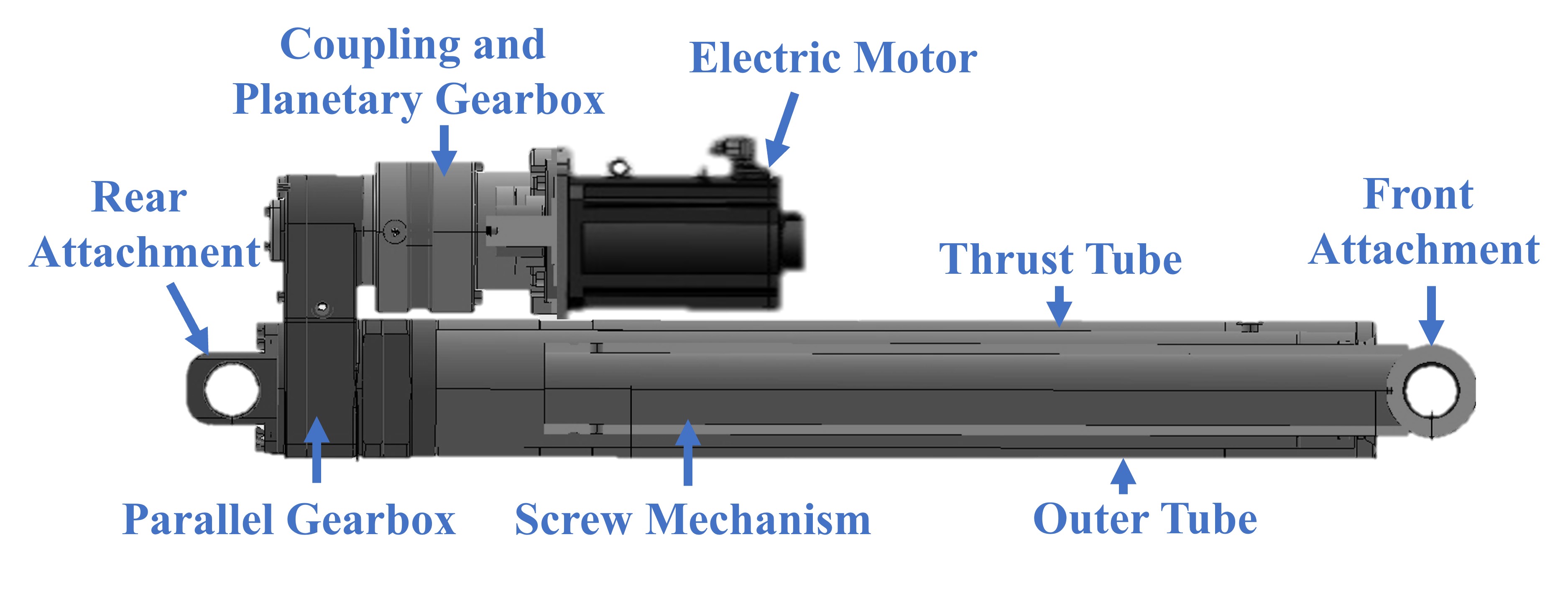}
	\caption{Schematic of the EMLA mechanism}
	\label{EMLA_structure}
\end{figure}

\begin{figure} [tbp]
    \centering
    \includegraphics[trim={0.0cm 0.0cm 0.0cm 0.0cm},clip,width=9.5cm]{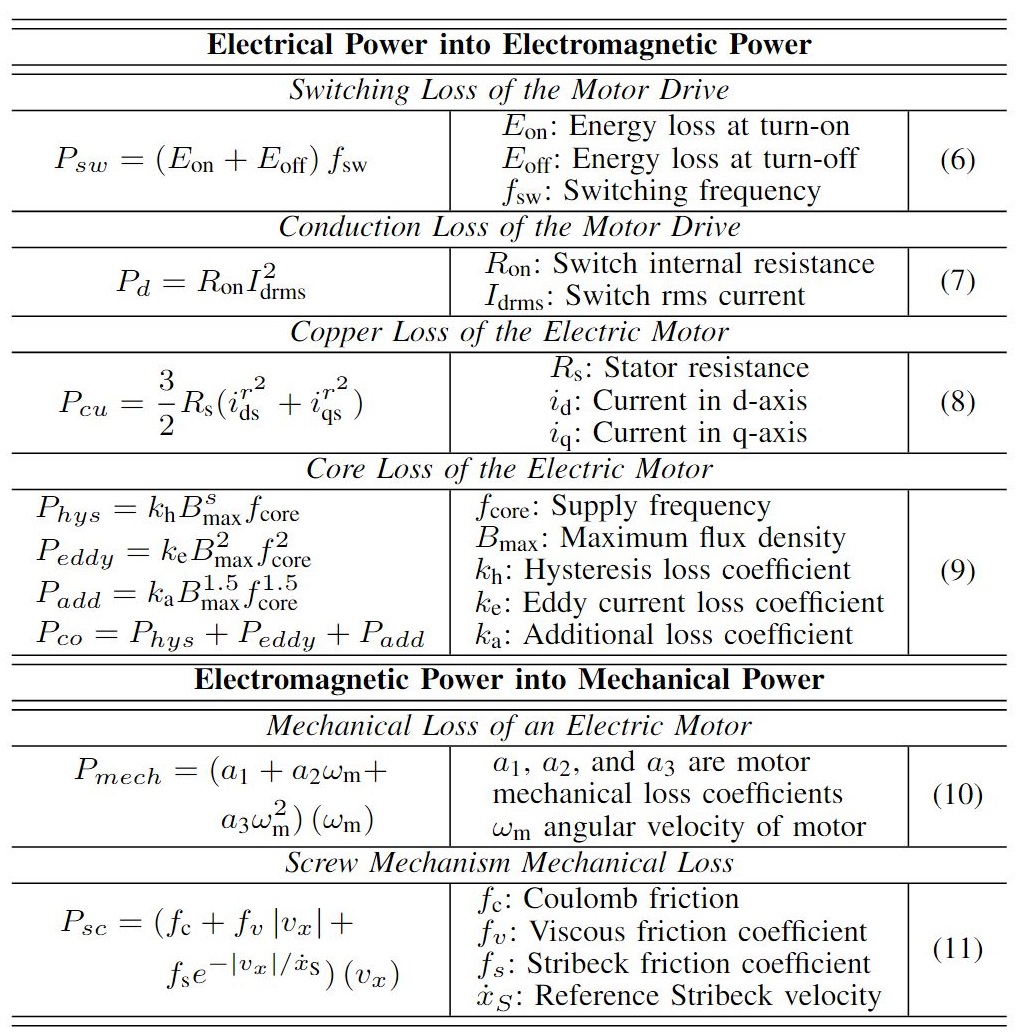}
    \caption{Power flow and dissipation sources in an EMLA}
    \label{power_losses}
\end{figure}

\begin{figure} [tbp]
	\centering
	\includegraphics[trim={0.0cm 0.0cm 0.0cm 0.0cm},clip,width=7.5cm]{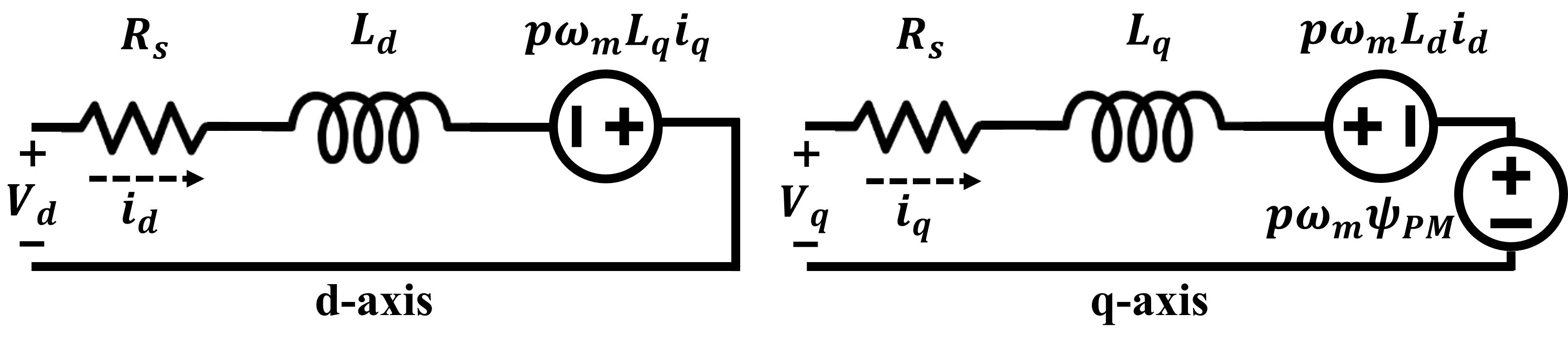}
	\caption{Equivalent circuits of PMSM in $dq$ reference frame}
	\label{dq_axis_circuit}
\end{figure}

In this section, the energy transformations occurring within an EMLA mechanism that incorporates an electric motor and gearbox are investigated. EMLAs, as depicted in Fig. \ref{EMLA_structure}, include a PMSM, which serves as the central component of the propulsion system, and it is coupled mechanically to the driving side of the gearbox to facilitate the transmission of rotary motion. Upon incorporating Park’s transformation into the voltage equations and converting them to the $dq$ reference frame, the equivalent circuits of the PMSM can be acquired, as illustrated in Fig. \ref{dq_axis_circuit}. In addition, the voltages of the PMSM within the $dq$ reference frame ($V_d$ and $V_q$) can be formulated as \eqref{equation:angle_transformation}-\eqref{equation:dq0voltage} \cite{9329162}:
\begin{equation}
\left\{
\begin{alignedat}{3}
&\theta_\gamma &&= \theta_m(t) \\
&\theta_\delta &&= \theta_m (t) - \frac{2\pi}{3}\\
&\theta_\epsilon &&=\theta_m (t) + \frac{2\pi}{3}
\end{alignedat}
\right.
\label{equation:angle_transformation}
\end{equation}
\begin{equation}
\hspace{-0.25cm} 
\begin{bmatrix}
V^r_{ds} \\
V^r_{qs} \\
V^r_0
\end{bmatrix}
= \frac{2}{3}
\begin{bmatrix}
\cos(\theta_\gamma) & \cos\left(\theta_\delta \right) & \cos\left(\theta_\epsilon \right) \\
-\sin(\theta_\gamma) & -\sin\left(\theta_\delta \right) & -\sin\left(\theta_\epsilon \right) \\
\frac{1}{2} & \frac{1}{2} & \frac{1}{2}
\end{bmatrix}
\begin{bmatrix}
V^s_a \\
V^s_b \\
V^s_c
\end{bmatrix}
\label{equation:park_transformation}
\end{equation}
\begin{equation}
\left\{
\begin{aligned}
V^r_{ds} &= R_s i^r_{ds} + L_d \frac{d i^r_{ds}}{dt} - p \omega_m L_q i^r_{qs} \\
V^r_{qs} &= R_s i^r_{qs} + L_q \frac{d i^r_{qs}}{dt} + p \omega_m i^r_{ds} L_d + p \omega_m \Psi_{P\!M}
\end{aligned}
\right.
\label{equation:dq0voltage}
\end{equation}
\hspace{-0.13cm}where $p$ represents the number of pole pairs; $L_d$ and $L_q$ are the motor inductance in the $d$- and $q$-axis, respectively; and ${\Psi_{PM}}$ signifies the flux linkage of the permanent magnet in the electric motor. It is worth mentioning the rest of the symbols are described in Table \ref{power_losses}. The general expression for electromagnetic torque of the PMSM can be derived as \eqref{equation:electromagnetictorque} \cite{xu2023permanent}:
\begin{equation}
\tau_{m}=\frac{3}{2} p i^r_{qs}\left[\Psi^r_{P\!M}+\left(L_d-L_q\right) i^r_{ds}\right],
\label{equation:electromagnetictorque}
\end{equation}

Within the EMLA, the rotational output of the driven gearbox integrates with a screw mechanism. This mechanism serves the function of converting rotary motion into linear motion, to drive linear load. Under the assumption of an ideal transmission system, the power transformation between the PMSM rotor and the load is governed by the fundamental relationships elucidated in \eqref{torque_force}.
\begin{equation}
\left\{
\begin{alignedat}{2}
&\tau_m &&= \frac{\rho}{2 \pi n} f_{\!x}\\
&\omega_m &&= \frac{2 \pi n}{\rho} v_{x}
\end{alignedat}
\right.
\label{torque_force}
\end{equation}
\begin{table}
\end{table}
\hspace{-0.13cm}where $f_{\!x}$ and $v_{x}$ are the force and linear velocity at the load side of the EMLA, respectively. Furthermore, $\rho$ is the screw lead and $n$ denotes the gear ratio, which is calculated by dividing the number of teeth on the driven side by the driving side. To facilitate performance evaluation of the EMLA, Table \ref{power_losses} presents analytical expressions for power losses occurring within the actuator, as detailed in equations (6)-(11).

\begin{figure} [tbp]
    \centering
    \includegraphics[trim={0.0cm 0.0cm 0.0cm 0.0cm},clip,width=8.5cm]{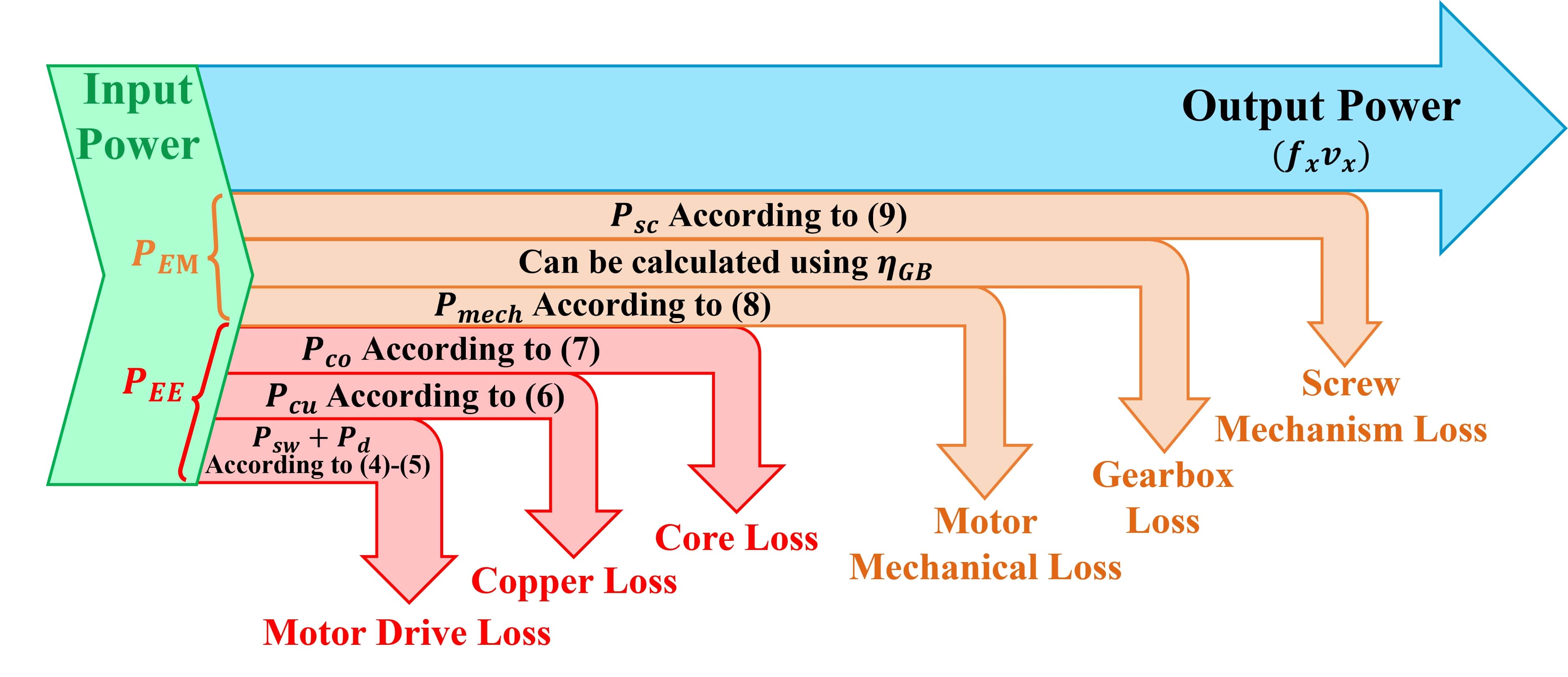}
    \caption{Power flow and dissipation sources in an EMLA}
    \label{powerflow}
\end{figure}

Fig. \ref{powerflow} visually depicts the power flow within the EMLA, highlighting the sources of power dissipation. By analyzing the energy conversion ratio of each component, we can quantify the power losses associated with the electric-to-electromagnetic ($P_{E\!E}$) and electromagnetic-to-mechanical ($P_{E\!M}$) conversions, as expressed in \eqref{EE_EM_Losses}. Subsequently, by employing \eqref{system_efficiency}, the overall efficiency of the EMLA can be calculated.
\setcounter{equation}{11}
\begin{equation}
\left\{
\begin{alignedat}{2}
&P_{E\!E} &&= P_{sw} + P_{d} + P_{cu} + P_{co}
\\
&P_{E\!M} &&= P_{mech} + P_{sc}
\end{alignedat}
\right.
\label{EE_EM_Losses}
\end{equation}
\begin{equation}
\eta_{\text{EMLA}}( f_{\!x}, v_{x} ) = \frac{f_{\!x} v_{x}}{f_{\!x} v_{x} + P_{E\!E} + P_{E\!M}}
\label{system_efficiency} 
\end{equation} 
\section{Dynamic and State Space Model of an EMLA}
\label{section:state_space_EMLA}
\begin{figure}[tbp] 
    \centering
    \includegraphics[trim={0.0cm 0.0cm 0.0cm 0.0cm},clip,width=8.5cm]{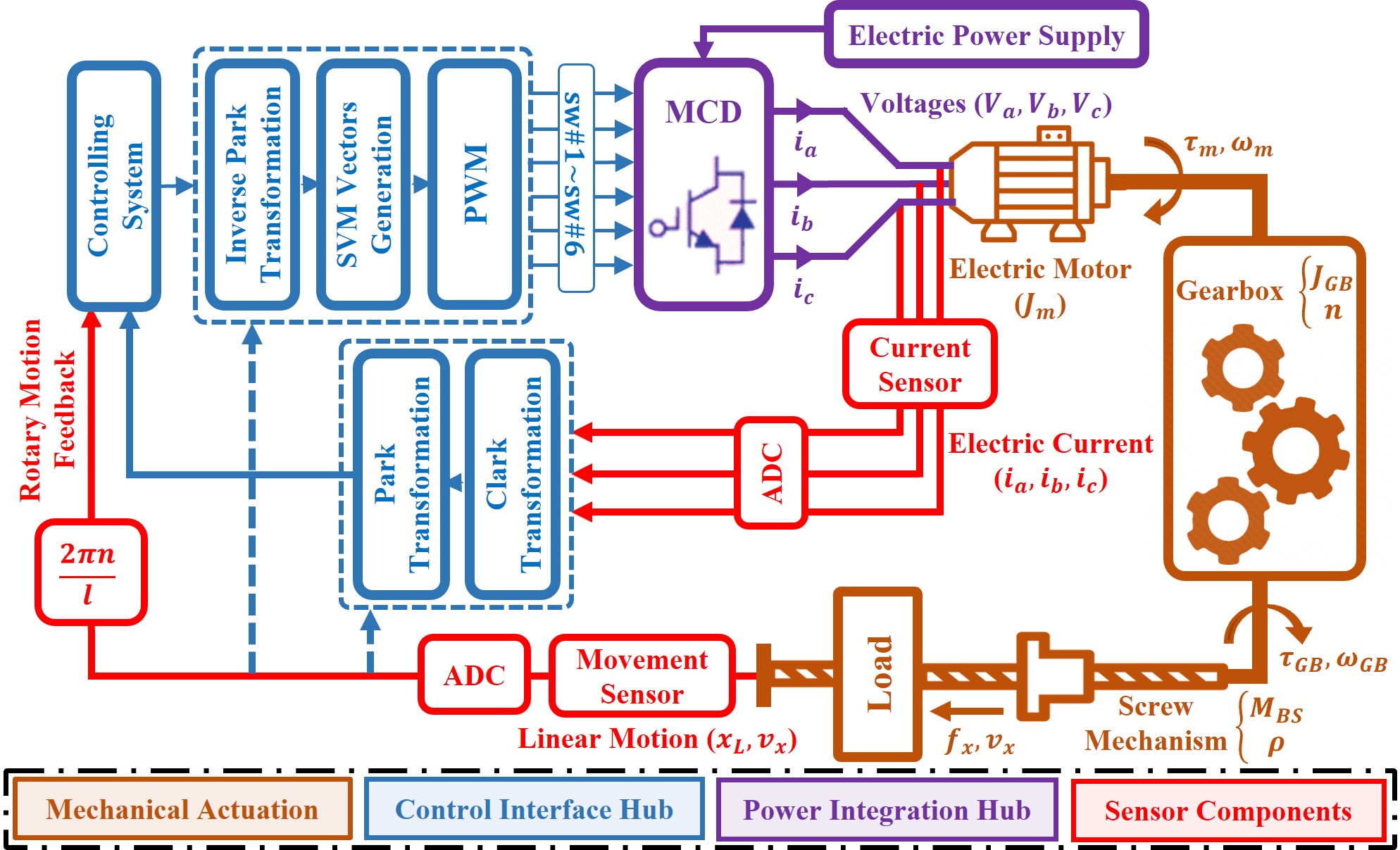}
    \caption{Decomposed schematic representation of the EMLA}
    \label{EMLA_schematic}
\end{figure}

Fig. \ref{EMLA_schematic} presents a schematic of the EMLA and the system's governing dynamics. To incorporate the mechanical aspect of the PMSM in the EMLA, the angular acceleration and angular velocity of the PMSM at rotor side are presented in \eqref{angularvelocity}:
\begin{equation}
\left\{
\begin{alignedat}{2}
&\frac{d^2 \theta_m}{dt^2} &&= \frac{1}{J_{eq}}(\tau_m - b_{eq} \frac{d \theta_m}{dt} - k_{eq} \theta_m - f_{eq} f_{\!x}) \\
&\frac{d \theta_m}{dt} &&= \omega_m
\end{alignedat}
\right.
\label{angularvelocity}
\end{equation}
where $J_{eq}$, $b_{eq}$, $k_{eq}$, and $f_{eq}$ denote the equivalent inertia, equivalent damping, equivalent stiffness, and load coefficient at the motor side, respectively, obtained as \eqref{equivalent_parameters}:
\begin{equation}
\left\{
\begin{alignedat}{3}
&J_{eq} &&= J_m + J_c + \frac{1}{n^2} J_{G\!B} + \frac{\rho ^ 2}{4 \pi^2 n^2} \left(M_{B\!S}+M_L\right)
\\
&b_{eq} &&= b_m + n b_{G\!B} + \frac{n \rho}{2 \pi} b_{B\!S}
\\
&k_{eq} &&= \frac{1}{k_{\tau 1}} + \frac{n^2}{k_{\tau 2}} + \frac{\left( \frac{2 \pi n}{\rho}\right)^2}{k_{L}}
\\
&f_{eq} &&= \frac{\rho}{2 \pi n}
\end{alignedat}
\right.
\label{equivalent_parameters}
\end{equation}

The system dynamics are characterized by inertia of the motor ($J_m$), coupling ($J_c$), and gearbox ($J_{GB}$). In addition, $M_{BS}$ and $M_L$ denote the mass of the screw mechanism and load. Also, damping effects are represented by the coefficients $b_m$, $b_{GB}$, and $b_{BS}$ for the motor, gearbox, and screw mechanism, respectively. Regarding system stiffness, $k_{\tau 1}$, $k_{\tau 2}$, and $k_{L}$ denote the torsional stiffness of the motor coupling, gear box, and screw shaft, respectively. Notably, the screw shaft’s torsional stiffness, as expressed in \eqref{linear_components_stiffness}, is a composite of the stiffness contributions from the thrust bearing ($k_{\text{bearing}}$), screw ($k_{\text{screw}}$), ball nut ($k_{\text{nut}}$), and thrust tube ($k_{\text{tube}}$).
\begin{equation}
k_L=\left(\frac{1}{k_{\text {bearing }}}+\frac{1}{k_{\text {screw }}}+\frac{1}{k_{\text {nut }}}+\frac{1}{k_{\text {tube }}}\right)^{-1}
\label{linear_components_stiffness}
\end{equation}
%

To achieve the purpose of this paper, dynamic equations are formulated based on the rotary motion of the EMLA at the motor side. Accordingly, \eqref{equation:dq0voltage}, \eqref{equation:electromagnetictorque}, and \eqref{angularvelocity}-\eqref{linear_components_stiffness} are employed to develop the state space model of the studied actuator. It should be noted that the state vector $\boldsymbol{x}\in\mathbb{R}^4$ and input vector $\boldsymbol{u}\in\mathbb{R}^2$ are defined in \eqref{state_space_2} for our case study.
\begin{equation}
\left\{
\begin{aligned}
\boldsymbol{\dot{x}} &= \mathcal{A}\boldsymbol{x}+\mathcal{B}\boldsymbol{u}+\boldsymbol{r}\\
\boldsymbol{x} &= \left[\begin{array}{llll}
\theta_m & \omega_m & i_q & i_d 
\end{array}\right]^T \\
\boldsymbol{u} &= \left[\begin{array}{ll}
V_q & V_d
\end{array}\right]^T
\end{aligned}
\right.
\label{state_space_2} 
\end{equation}

\begin{figure}[tbp]
    \centering
    \includegraphics[trim={0.0cm 0.0cm 0.0cm 0.0cm},clip,width=9cm]{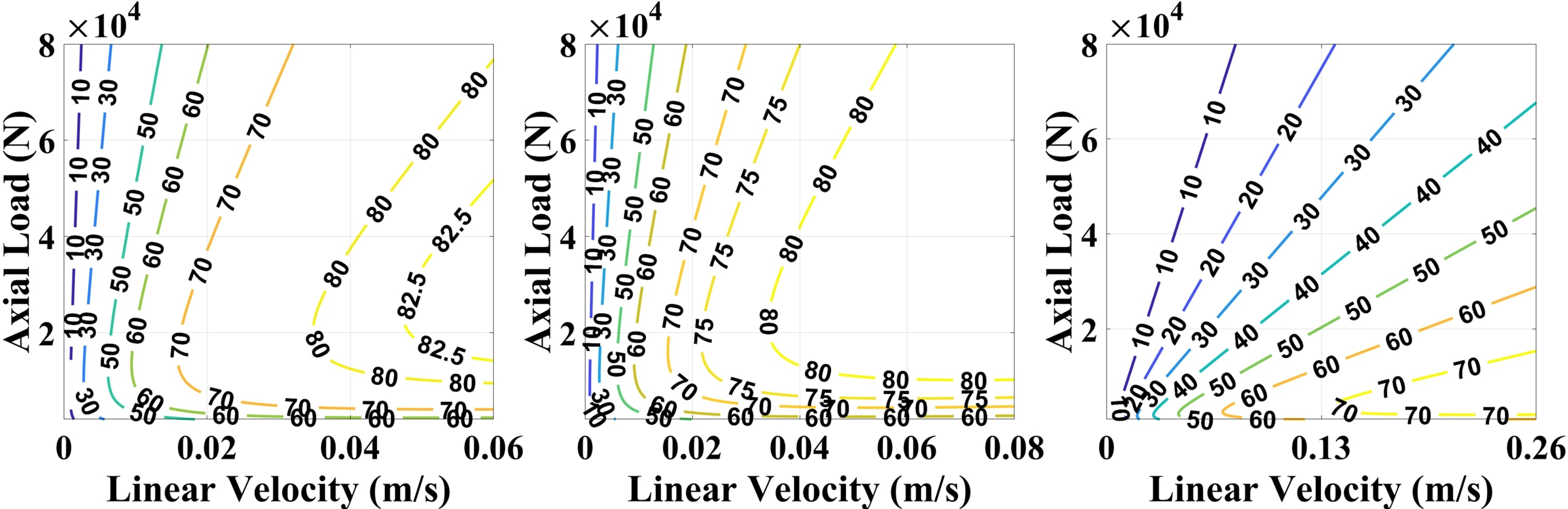}
    \caption{Efficiency maps of the EMLAs in Table \ref{EMLA_Table}: lift (left), tilt (middle), and telescope (right)}
    \label{Efficiency_3DOF}
\end{figure}

\begin{figure}[tbp]
    \centering
    \includegraphics[trim={0.0cm 0.0cm 0.0cm 0.0cm},clip,width=8.5cm]{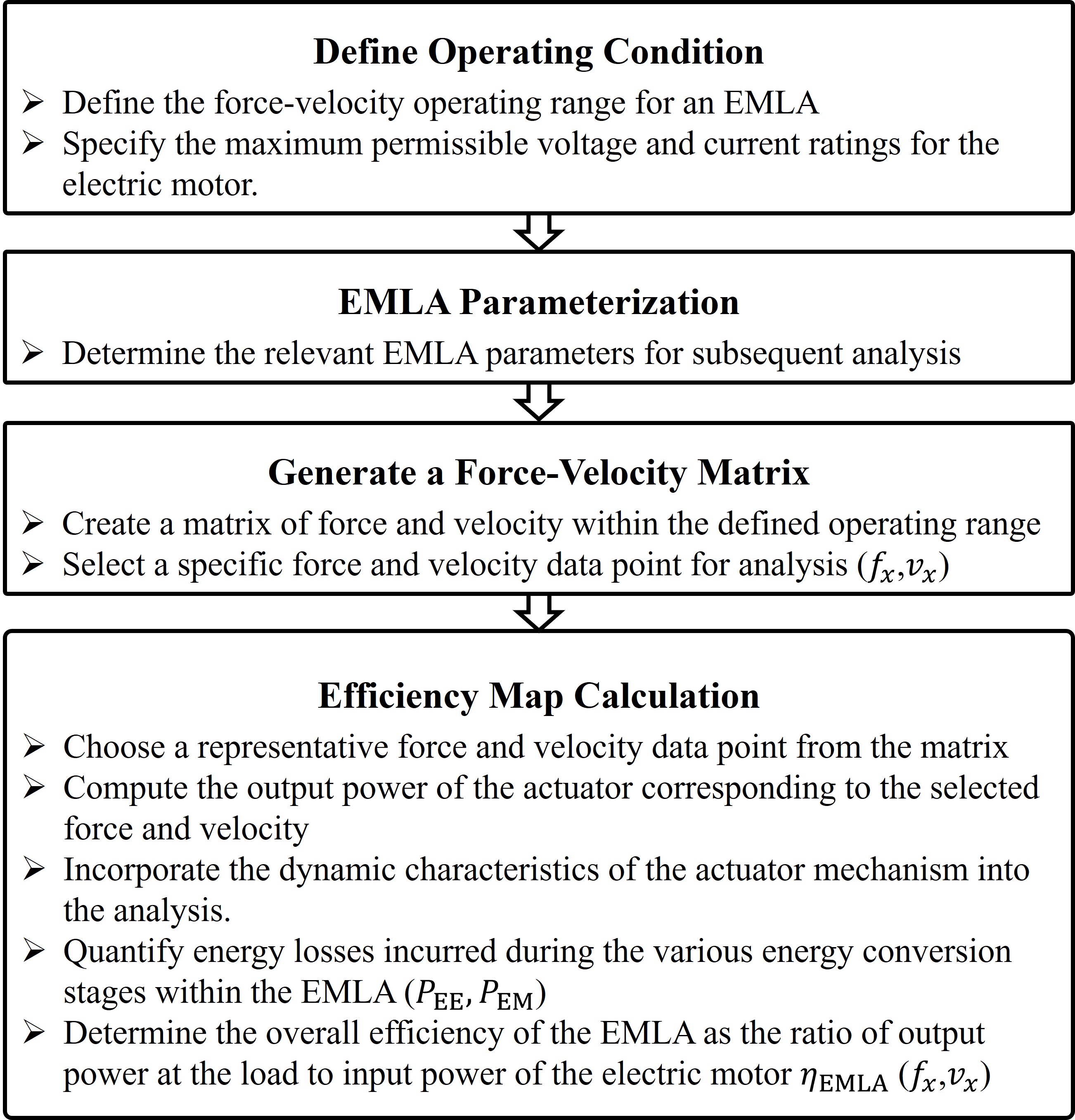}
    \caption{Process for efficiency evaluation of an EMLA}
    \label{Efficiency_Flowchart}
\end{figure}

By employing linearization and incorporating the approximations of $i_d \omega_m$ and $i_q \omega_m$ into \eqref{state_space_2}, the values of $\mathcal{A}$, $\mathcal{B}$, and $\boldsymbol{r}$ can be determined for the state space model, as in \eqref{matrix_A}-\eqref{matrix_R}:

\begin{equation}
\hspace{-0.25cm} 
\mathcal{A} = \begin{bmatrix}
    -\frac{R_s}{L_d} & p \alpha^{-1} x_2^0 & p \alpha^{-1} x_3^0 & 0 \\
    -p \alpha x_2^0 & -\frac{R_s}{L_q} & -\left(p \alpha x_4^0 + \frac{p \Psi_{P\!M}}{L_q}\right) & 0 \\
    \frac{\Delta L \hspace{0.5mm} x_3^0}{\beta} & \frac{\Delta L \hspace{0.5mm} x_4^0 + \Psi_{P\!M}}{\beta} & -\frac{b_{e\!q}}{J_{e\!q}} & 1 \\
    0 & 0 & 1 & 0
\end{bmatrix}
\label{matrix_A}
\end{equation}
\begin{equation}
\text{where \hspace{3mm}}
\left\{
\begin{alignedat}{3}
&\alpha &&= \frac{L_d}{L_q} \\
&\beta  &&= \frac{2 J_{eq}}{3 p} \\
&\Delta L &&= L_d - L_q
\end{alignedat}
\right.
\text{and \hspace{3mm}} \mathcal{B} = \begin{bmatrix}
    \frac{1}{L_d} & 0 \\
    0 & \frac{1}{L_q} 
\end{bmatrix}
\label{matrix_B}
\end{equation}
\begin{equation}
\boldsymbol{r} = \begin{bmatrix}
    -\frac{p}{\alpha} x_2^0 x_3^0 \\
    -p \alpha x_2^0 x_4^0 \\
    -\frac{\Delta L}{\beta}  x_3^0 x_4^0 - f_{\!{eq}} f_{\!x}\\
    0
\end{bmatrix}
\label{matrix_R}
\end{equation}
where $x_2^0$, $x_3^0$, and $x_4^0$ are the operating points of $\omega_m$, $i_q$, and $i_d$, respectively. Leveraging the analytical expressions for power losses presented in (6)-(11) and capturing dynamic behaviour in \eqref{state_space_2}-\eqref{matrix_R}, the efficiency of the selected EMLAs in Table \ref{EMLA_Table} are computed as a function of the axial force and linear velocity using \eqref{system_efficiency} and mapped in Fig. \ref{Efficiency_3DOF}. The detailed methodology for this calculation is outlined in the flowchart depicted in Fig. \ref{Efficiency_Flowchart}.
\begin{table}[tbp]
    \centering
    \caption{Overview of selected EMLAs implemented in studied robotic manipulator joints}
    \begin{tabular}{c|c|c|c}
        \hline
        \hline
        \textbf{Joint} & \textbf{Linear Unit} & \textbf{Motor Model/Manufacturer} & \textbf{Power}\\
        \hline
        Lift & SRSA-S-4805 & MCS14H28/Lenze & 6.0kW\\
        \hline
        Tilt & SRSA-S-3910 & MCS14H32/Lenze & 4.7kW\\
        \hline
        Telescope & CASM-100-BA & 1FK7064/Siemens & 2.5kW\\
        \hline
        \hline
    \end{tabular}
    \label{EMLA_Table}
\end{table}
\section{Kinematic and Dynamic Analysis of a Heavy-Duty Manipulator Structure}
\label{kinematics_and_dynamic}
\begin{figure}[b]
    \centering
    \includegraphics[trim={0.0cm 0.0cm 0.0cm 0.0cm},clip,width=9cm]{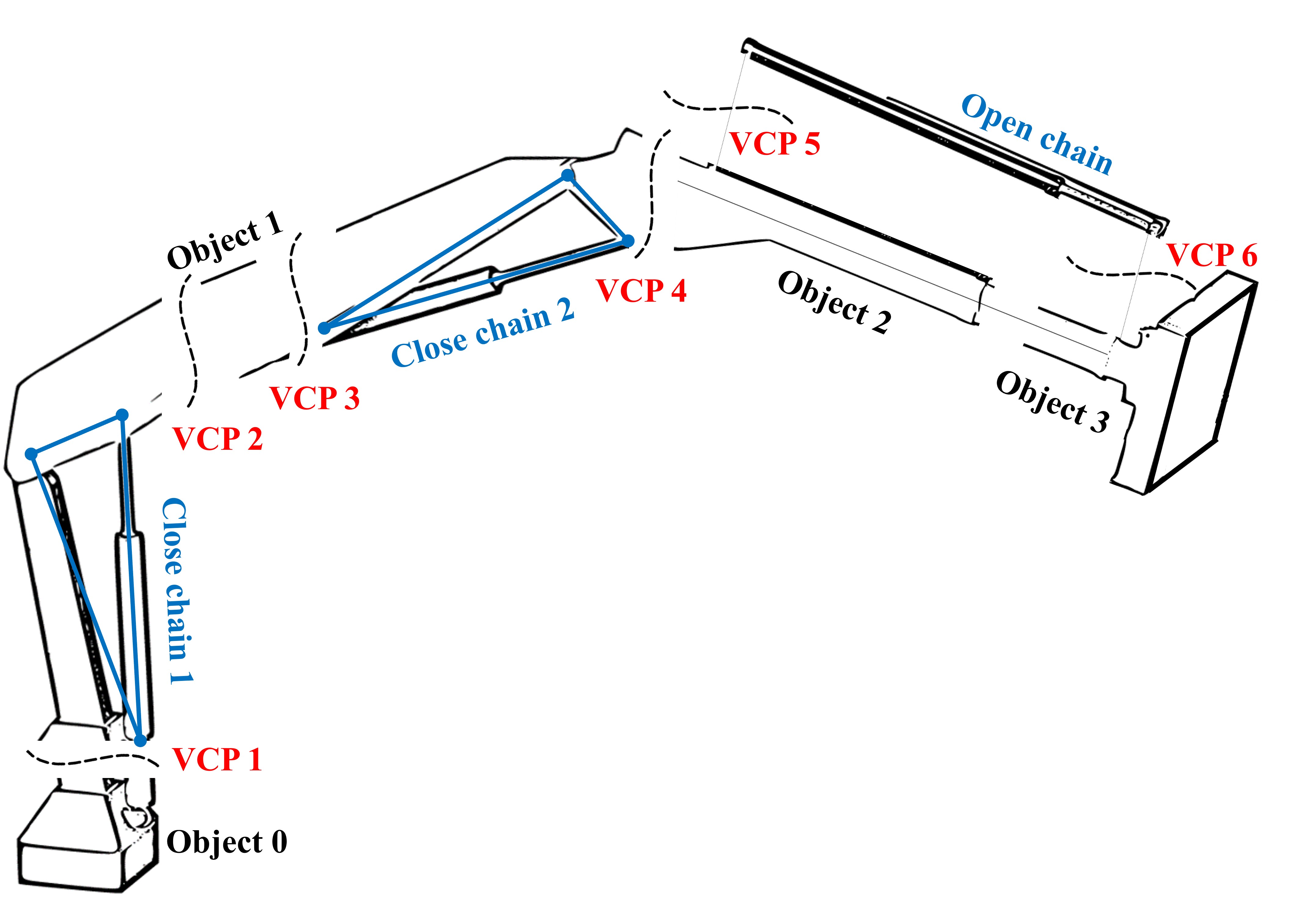}
    \caption{Virtually decomposed subsystems of the studied heavy-duty manipulator}
    \label{fig:decomposed}
\end{figure}

\begin{figure}[tbp]
    \centering
    \includegraphics[trim={0.0cm 0.0cm 0.0cm 0.0cm},clip,width=9cm]{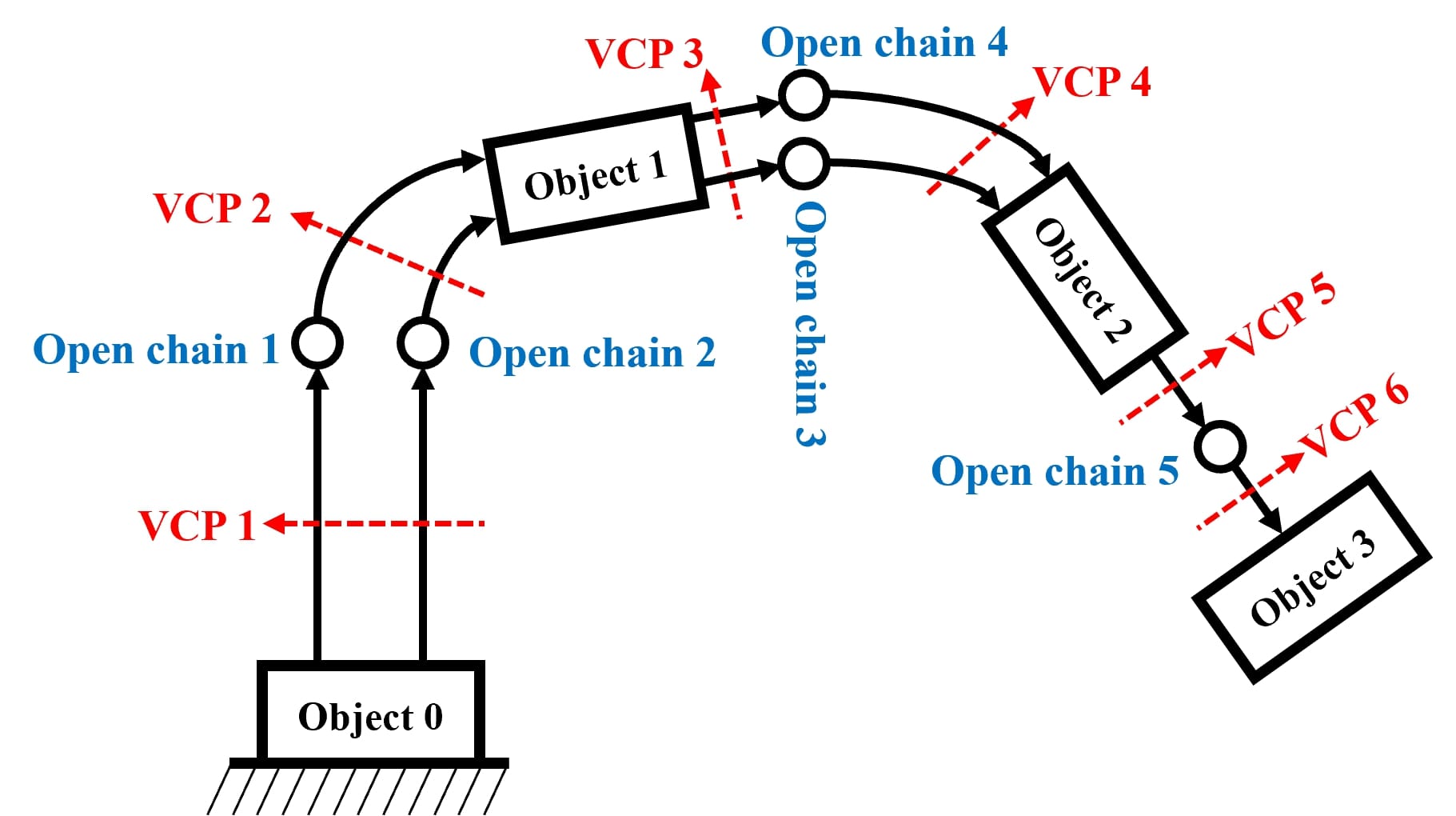}
    \caption{Oriented graph of the studied heavy-duty manipulator}
    \label{fig:Oriented_graph}
\end{figure}

In this section, the equations of motion governing the heavy-duty manipulator are meticulously derived. To achieve this, we employ the virtual decomposition control (VDC) method to accurately model and analyze the complex structure of a serial parallel heavy-duty manipulator. VDC simplifies the modeling process by decoupling joint dynamics, enhancing computational efficiency and accuracy \cite{zhu2010virtual}. It effectively addresses challenges such as unknown friction and gravitational torque, ensuring precise analysis of the system \cite{7852495}. Additionally, VDC’s integration into adaptive control schemes enables robust interaction force control, even in unknown environments, while its low computational complexity meets real-time control requirements \cite{887620}. By treating each subsystem separately, VDC simplifies the overall analysis and ensures the robustness of the derived equations \cite{7130666}. As illustrated in Fig. \ref{fig:decomposed}, the entire complex system is partitioned into three objects, each subjected to independent analysis. This decomposition effectively addresses the intricacies inherent in the structure of the manipulator \cite{brahmi2016adaptive,luna2016virtual,10000105}. The system under consideration features a sophisticated combination of components, including open- and close-chains (serial-parallel joint) configurations. By leveraging the VDC method, each subsystem, whether representing a rigid body or an actuator, is treated individually, thereby simplifying the overall analysis \cite{9891804}. A simple oriented graph of the studied heavy-duty manipulator is visualized in Fig. \ref{fig:Oriented_graph}
\subsection{Fundamental Concepts of VDC}
\label{fundamental_vdc}
Each rigid body in the analysis can have an attached frame, referred to as $\{\mathbf{A}\}$, where $\overrightarrow{\mathbf{f}_{\mathbf{A}}}$ and $\overrightarrow{\mathbf{m}_{\mathbf{A}}}$ are the force and moment vectors applied to it, while $\overrightarrow{\mathbf{v}_{\mathbf{A}}}$ and $\overrightarrow{\boldsymbol{\omega}_{\mathbf{A}}}$ are the linear and angular velocity vectors, as defined in \eqref{eq:vdc_vectors} \cite{zhu2010virtual}.

\begin{equation}
\left\{
\begin{aligned}
\overrightarrow{\mathbf{f}}_{\mathbf{A}}&=\{\mathbf{A}\}^{\mathbf{A}} \mathbf{f}\\
\overrightarrow{\mathbf{m}}_{\mathbf{A}}&=\{\mathbf{A}\}^{\mathbf{A}} \mathbf{m}\\
\overrightarrow{\mathbf{v}}_{\mathbf{A}}&=\{\mathbf{A}\}^{\mathbf{A}} \mathbf{v}\\
\overrightarrow{\boldsymbol{\omega}}_{\mathbf{A}}&=\{\mathbf{A}\}^{\mathbf{A}} \boldsymbol{\omega}
\end{aligned}
\right.
\label{eq:vdc_vectors}
\end{equation}

Accordingly, considering $\mathbf{{}^{A}\boldsymbol{f}}$ $\in \mathbb{R}^3$, $\mathbf{{}^{A}\boldsymbol{m}}$ $\in \mathbb{R}^3$, $\mathbf{{}^{A}\boldsymbol{v}}$ $\in \mathbb{R}^3$, and $\mathbf{{}^{A}\boldsymbol{\omega}}$ $\in \mathbb{R}^3$, the linear/angular velocity vector and force/moment vector of frame $\{\mathbf{A}\}$ can be defined as \eqref{eq:linear_angular_V_F}.
\begin{equation}
\left\{
\begin{aligned}
\mathbf{{}^{A}}{\boldmath{V}} &\stackrel{\textit{def}}{=}\left[\begin{array}{c}
\mathbf{{}^{A}\boldsymbol{v}}  \\
\mathbf{{}^{A}\boldsymbol{\omega}} 
\end{array}\right] &\in \mathbb{R}^6 \\
\mathbf{{}^{A}}{\!\boldmath{F}} &\stackrel{\textit{def}}{=}\left[\begin{array}{c}
\mathbf{{}^{A}\boldsymbol{f}} \\
\mathbf{{}^{A}\boldsymbol{m}}
\end{array}\right] &\in \mathbb{R}^6 
\end{aligned}
\right.
\label{eq:linear_angular_V_F}
\end{equation}

Given two frames, $\{\mathbf{A}\}$ and $\{\mathbf{B}\}$, attached to a common rigid body in motion and subjected to physical force and moment vectors, the transformation matrix that converts force/moment and linear/angular velocity vectors between these two frames can be described by \eqref{eq:duality}.
\begin{equation}
\left\{
\begin{aligned}
\mathbf{{}^{B}}{\boldmath{V}} & ={ }^{\mathbf{A}} \mathbf{U}_{\mathbf{B}}^T \mathbf{{}^{A}}{\boldmath{V}} \\
\mathbf{{}^{A}}{\!\boldmath{F}} & ={ }^{\mathbf{A}} \mathbf{U}_{\mathbf{B}}{ }^{\mathbf{B}} {\!\boldmath{F}}
\end{aligned}
\right.
\label{eq:duality}
\end{equation}
where ${ }^{\mathbf{A}} \mathbf{U}_{\mathbf{B}} \in \mathbb{R}^{6 \times 6}$ can be obtained as \eqref{eq:transformation_matrix}.
\begin{equation}
{ }^{\mathbf{A}} \mathbf{U}_{\mathbf{B}}=\left[\begin{array}{cc}
{ }^{\mathbf{A}} \mathbf{R}_{\mathbf{B}} & \mathbf{0}_{3 \times 3} \\
\left({ }^{\mathbf{A}} \mathbf{r}_{\mathbf{A B}} \times\right) { }^{\mathbf{A}} \mathbf{R}_{\mathbf{B}} & { }^{\mathbf{A}} \mathbf{R}_{\mathbf{B}}
\end{array}\right] 
\label{eq:transformation_matrix}
\end{equation}

It is noteworthy that ${ }^{\mathbf{A}} \mathbf{R}_{\mathbf{B}} \in \mathbb{R}^{3 \times 3}$ in $SO(3)$ represents the rotation matrix between frame $\{\mathbf{A}\}$ and frame $\{\mathbf{B}\}$ \cite{wen1991attitude}. Additionally, ${ }^{\mathbf{A}} \mathbf{r}_{\mathbf{A B}} \in \mathbb{R}^{3 \times 3}$ is a skew-symmetric matrix operator that denotes the vector from the origin of frame $\{\mathbf{A}\}$ to the origin of frame $\{\mathbf{B}\}$, expressed in frame $\{\mathbf{A}\}$ and can be obtained as \eqref{eq:skew_symmetric_matrix}. 
\begin{equation}
\left({ }^{\mathbf{A}} \mathbf{r}_{\mathbf{A B}} \times\right)=\left(\begin{array}{ccc}
0 & -r_{\mathrm{z}} & r_{\mathrm{y}} \\
r_{\mathrm{Z}} & 0 & -r_{\mathrm{x}} \\
-r_{\mathrm{y}} & r_{\mathrm{x}} & 0
\end{array}\right)
\label{eq:skew_symmetric_matrix}
\end{equation}
where $r_x$, $r_y$, and $r_z$ represent the distances from the origin of frame $\{\mathbf{A}\}$ to the origin of frame $\{\mathbf{B}\}$ along the $x$-, $y$-, and $z$-axes of frame $\{\mathbf{A}\}$, respectively. Consider the net force and moment vectors acting on the rigid body with frame $\{\mathbf{A}\}$ attached, denoted as ${}^{\mathbf{A}} \boldsymbol{F}^* \in \mathbb{R}^6$, which can be computed based on its dynamics as \eqref{eq:net_force}.
\begin{equation}
\mathbf{{}^{A}}{\!\hat{\boldsymbol{F}}} = \mathbf{M}_{\mathbf{A}} \frac{\textit{d}}{\textit{dt}}\! \left({ }^{\mathbf{A}} V\right) + \mathbf{C}_{\mathbf{A}} \! \left({ }^{\mathbf{A}} \omega\right)^{\mathbf{A}}\! V + \mathbf{G}_{\mathbf{A}}
\label{eq:net_force}
\end{equation}
where $\mathbf{M}_{\mathbf{A}} \in \mathbb{R}^{6 \times 6}$ represents the mass matrix, $\mathbf{C}_{\mathbf{A}} \! \left({ }^{\mathbf{A}} \omega\right) \in \mathbb{R}^{6 \times 6}$ denotes the matrix of Coriolis and centrifugal terms, and $\mathbf{G}_{\mathbf{A}} \in \mathbb{R}^{6}$ corresponds to the gravity terms (see \eqref{eq:mass_matrix}-\eqref{gravity} in the Appendix for further details). The linear parametrization expression of \eqref{eq:net_force} for the mentioned rigid body can be expressed as \eqref{eq:linear_net_force} where ${ }^{\mathbf{A}} V_r \in \mathbb{R}^{6}$ is the required vector of ${ }^{\mathbf{A}} V \in \mathbb{R}^{6}$.
\begin{equation}
\mathbf{Y}_{\mathbf{A}} \boldsymbol{\theta}_{\mathbf{A}} \stackrel{\textit { def }}{=} \mathbf{M}_{\mathbf{A}} \frac{\textit{d}}{\textit{dt}}\! \left({ }^{\mathbf{A}} V_r\right) + \mathbf{C}_{\mathbf{A}} \! \left({ }^{\mathbf{A}} \omega\right)^{\mathbf{A}}\! V_r + \mathbf{G}_{\mathbf{A}}
\label{eq:linear_net_force}
\end{equation}
where $\mathbf{Y}_{\mathbf{A}} \in \mathbb{R}^{6 \times 13}$ is the regressor matrix and $\boldsymbol{\theta}_{\mathbf{A}} \in \mathbb{R}^{13}$ is the parameter vector.
\subsection{Mathematical Modeling of Heavy-Duty Manipulator}
\label{mathematical_modeling_HDMM}
In this section, the equations of motion for the 3-DoF heavy-duty manipulator are developed. As the VDC method utilizes the Newton--Euler (NE) approach, it necessitates computing forward velocity and backpropagating forces to derive the dynamic equations.
The first step in the VDC approach is to virtually decompose the original system into subsystems, such as objects and open chains, by introducing conceptual virtual cutting points (VCPs). The entire system is then represented as a simple directed graph, where each subsystem corresponds to a node and each VCP corresponds to a directed edge that defines the force reference direction. A VCP acts as a driving VCP for one subsystem from which the force/moment vector is exerted, with the directed edge pointing away, and as a driven VCP for another subsystem to which the force/moment vector is exerted, with the directed edge pointing toward. This graph effectively visualizes the dynamic interactions between subsystems after virtual decomposition, as shown in Fig \ref{fig:Oriented_graph}.
\subsubsection{Kinematics of the 3-DoF heavy-duty manipulator}
A linearly actuated parallel mechanism, as depicted in Fig. \ref{fig:parallel_mechanism}, comprises a closed kinematic chain featuring four joints: three passive revolute joints and one prismatic actuated joint. To support the VCP approach in a closed kinematic loop analysis, constraints must be introduced. These constraints involve using analytic loop--closure functions to relate the joint angle of each revolute segment to the corresponding piston displacement and the two closed-chain angles $q_{j1}$ and $q_{j2}$, based on known lengths. Referring to Fig. \ref{fig:parallel_mechanism} and considering the notation $j=1,2$ that represents the first and second closed chains, these relations are outlined as \eqref{eq:kinematics_constraints}:
\begin{figure}[btp]
    \centering
    \includegraphics[trim={0.0cm 0.0cm 0.0cm 0.0cm},clip,width=5.5cm]{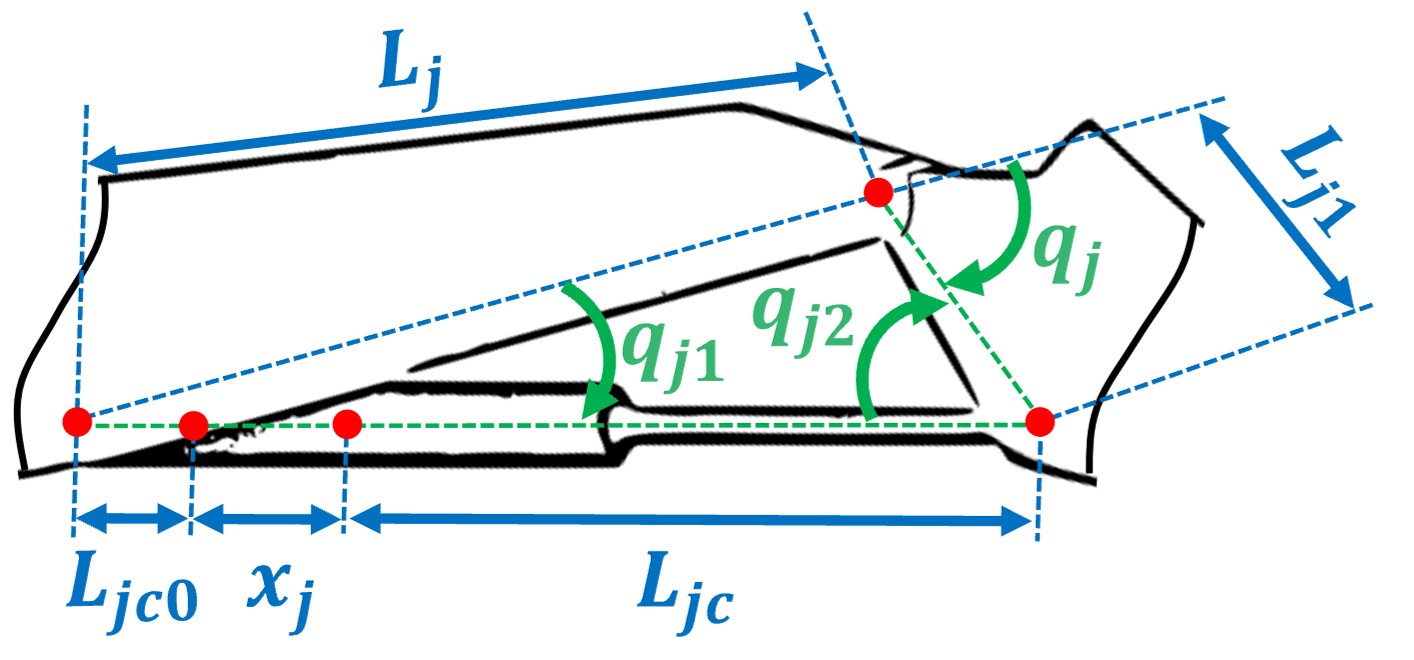}
    \caption{1-DoF parallel kinematic mechanism featuring three passive revolute joints and a single prismatic actuator.}
    \label{fig:parallel_mechanism}
\end{figure}
\begin{equation}
\left\{
\begin{aligned}
q_j & =-\arccos \left(\frac{\left(x_j+x_{j0}\right)^2-L_j^2-L_{j1}^2}{-2 L_j L_{j1}}\right) \\
q_{j1} & =-\arccos \left(\frac{L_{j1}^2-\left(x_j+x_{j0}\right)^2-L_j^2}{-2\left(x_j+x_{j0}\right) L_j}\right) \\
q_{j2} & =-\arccos \left(\frac{L_j^2-\left(x_j+x_{j0}\right)^2-L_{j1}^2}{-2\left(x_j+x_{j0}\right) L_{j1}}\right)
\end{aligned}
\right.
\label{eq:kinematics_constraints}
\end{equation}
where $L_j$ and $L_{j1}$ are the link lengths of open chains between the frames, while $x_{j0}$ is the effective length of the EMLA at the zero stroke position:
\begin{equation}
\left\{
\begin{aligned}
L_j & > 0 \\
L_{j1} & > 0 \\
x_{j0} & = L_{jc} + L_{jc0}
\end{aligned}
\right.
\label{eq:link_lengths}
\end{equation}

\begin{figure*}[tbp]
    \centering
    \includegraphics[trim={0.0cm 0.0cm 0.0cm 0.0cm},clip,width=15cm]{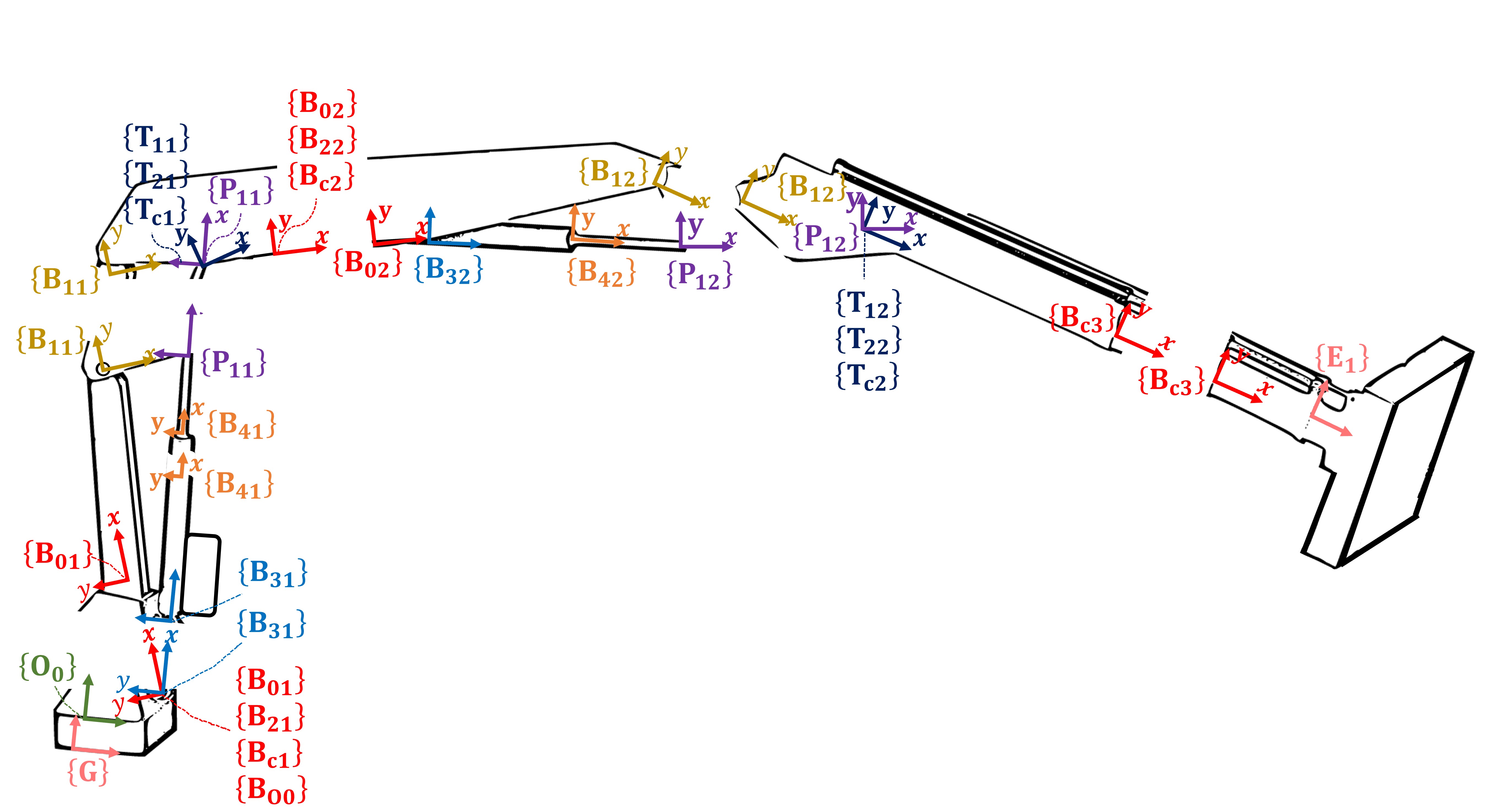}
    \caption{VDC frames of the 3-DoF heavy-duty manipulator}
    \label{fig:VDC_frames}
\end{figure*}

After decomposing the system into subsystems based on the designated VCPs (refer to Fig. \ref{fig:decomposed}) and the assigned VDC frames (refer to Fig. \ref{fig:VDC_frames}), the velocity of object 0 can be expressed as a function of the ground velocity, ${}^{\mathbf{G}} \boldsymbol{V}$:
\begin{equation}
{}^{\mathbf{O}_{\mathrm{0}} } \boldsymbol{V}={ }^{\mathbf{G}} \mathbf{U}_{\mathbf{O}_{ \mathrm{0}}}^T {}^{\mathbf{G}} \boldsymbol{V}
\end{equation}

In the next step, any existing closed chain structures are also virtually broken down into open chains to ensures the entire system only consists of objects and open chains. This allows us to express the linear and angular velocities within the open kinematic chains $\mathrm{1 j}$ and $\mathrm{2 j}$ lower serial chain, as: 
\begin{equation}
\left\{
\begin{aligned}
{}^{\mathbf{B}_{3 \mathrm{j}} } \boldsymbol{V}&=\mathbf{z}_\tau \dot{q}_{j1}+{ }^{\mathbf{B}_{2 \mathrm{j}}} \mathbf{U}_{\mathbf{B}_{3 \mathrm{j}}}^T {}^{\mathbf{B}_{2 \mathrm{j}}} \boldsymbol{V} \\
{}^{\mathbf{B}_{4 \mathrm{j}} } \boldsymbol{V}&=\mathbf{x}_f \dot{x}_{j}+{ }^{\mathbf{B}_{3 \mathrm{j}}} \mathbf{U}_{\mathbf{B}_{4 \mathrm{j}}}^T {}^{\mathbf{B}_{3 \mathrm{j}}} \boldsymbol{V} \\
{}^{\mathbf{T}_{2 \mathrm{j}} } \boldsymbol{V}&=\mathbf{z}_\tau \dot{q}_{j2}+{ }^{\mathbf{B}_{4 \mathrm{j}}} \mathbf{U}_{\mathbf{T}_{2 \mathrm{j}}}^T {}^{\mathbf{B}_{4 \mathrm{j}}} \boldsymbol{V}
\end{aligned}
\right.
\end{equation}

Those of the upper serial chain's open kinematic chains $\mathrm{1 j}$ and $\mathrm{2 j}$ can also be expressed as:
\begin{equation}
\left\{
\begin{aligned}
{}^{\mathbf{B}_{c \mathrm{1}} } \boldsymbol{V}&={}^{\mathbf{O}_{0}} \mathbf{U}_{\mathbf{B}_{c \mathrm{1}}}^T {}^{\mathbf{O}_{0}} \boldsymbol{V}\\
{}^{\mathbf{B}_{1 \mathrm{j}} } \boldsymbol{V}&=\mathbf{z}_\tau \dot{q}_j+{ }^{\mathbf{B}_{0 \mathrm{j}}} \mathbf{U}_{\mathbf{B}_{1 \mathrm{j}}}^T {}^{\mathbf{B}_{0 \mathrm{j}}}\boldsymbol{V} \\
{}^{\mathbf{T}_{1 \mathrm{j}} } \boldsymbol{V}&={}^{\mathbf{B}_{1 \mathrm{j}}} \mathbf{U}_{\mathbf{T}_{1 \mathrm{j}}}^T {}^{\mathbf{B}_{1 \mathrm{j}}} \boldsymbol{V}
\end{aligned}
\right.
\end{equation}

The linear and angular velocities of object 1, as measured and described in frame $\{\mathbf{B_{c2}}\}$, can be determined as:
\begin{align}
    {}^{\mathbf{B}_{ \mathrm{c 2}} } \boldsymbol{V} &= {}^{\mathbf{T}_{ \mathrm{c 1}}} \mathbf{U}_{\mathbf{B}_{ \mathrm{c 2}}}^T {}^{\mathbf{T}_{ \mathrm{c 1}}} \boldsymbol{V}
\end{align}

In reference to Fig. \ref{fig:VDC_frames}, the linear and angular velocities at the driving VCP of the closed kinematic chains are:
\begin{equation}
\left\{
\begin{aligned}
{}^{\mathbf{T}_{c \mathrm{j}} } \boldsymbol{V}={}^{\mathbf{T}_{1 \mathrm{j}} } \boldsymbol{V}={}^{\mathbf{T}_{2 \mathrm{j}} } \boldsymbol{V} \\
{}^{\mathbf{B}_{c \mathrm{1}} } \boldsymbol{V}={}^{\mathbf{B}_{0 \mathrm{j}} } \boldsymbol{V}={}^{\mathbf{B}_{2 \mathrm{j}} } \boldsymbol{V}
\end{aligned}
\right.
\end{equation}

The linear and angular velocities of object 2, as observed and defined in the reference frame $\{\mathbf{E_{1}}\}$, are given by:%
\begin{equation}
\left\{
\begin{aligned}
    {}^{\mathbf{B}_{c \mathrm{3}} } \boldsymbol{V} &= {}^{\mathbf{T}_{c2}} \mathbf{U}_{\mathbf{B}_{c \mathrm{3}}}^T {}^{\mathbf{T}_{c2}} \boldsymbol{V}\\
    {}^{\mathbf{E}_{ \mathrm{1}} } \boldsymbol{V} &= {}^{\mathbf{B}_{c3}} \mathbf{U}_{\mathbf{E}_{\mathrm{1}}}^T {}^{\mathbf{B}_{c3}} \boldsymbol{V}    
\end{aligned}
\right.
\end{equation}
%

%
\subsubsection{Dyanmics of the 3-DoF heavy-duty manipulator}
This part derives the spatial force vectors acting on each rigid body within the system. The net forces obtained in \eqref{eq:net_force} are calculated for each rigid body by substituting the appropriate frame for $\{\mathbf{A}\}$.
\begin{equation}
\left\{
\begin{aligned}
    {}^{\mathbf{B}_{0 \mathrm{j}} } \boldsymbol{F}&=
    {}^{\mathbf{B}_{0 \mathrm{j}} } \hat{\boldsymbol{F}} +
    {}^{\mathbf{B}_{0j}} \mathbf{U}_{\mathbf{B}_{1 \mathrm{j}}} {}^{\mathbf{B}_\mathrm{1j}} \boldsymbol{F} \\
    {}^{\mathbf{B}_{1 \mathrm{j}} } \boldsymbol{F}&=
    {}^{\mathbf{B}_{1 \mathrm{j}} } \hat{\boldsymbol{F}} +
    {}^{\mathbf{B}_{1j}} \mathbf{U}_{\mathbf{B}_{c \mathrm{j+1}}} {}^{\mathbf{B}_\mathrm{cj+1}} \boldsymbol{F} -
    {}^{\mathbf{B}_{1j}} \mathbf{U}_{\mathbf{P}_{1 \mathrm{j}}} {}^{\mathbf{P}_\mathrm{1j}} \boldsymbol{F} \\
    {}^{\mathbf{B}_{2 \mathrm{j}} } \boldsymbol{F}&=
    {}^{\mathbf{B}_{2j}} \mathbf{U}_{\mathbf{B}_{\mathrm{3j}}} {}^{\mathbf{B}_{3j}} \boldsymbol{F}\\
    {}^{\mathbf{B}_{3 \mathrm{j}} } \boldsymbol{F}&=
    {}^{\mathbf{B}_{3 \mathrm{j}}} \hat{\boldsymbol{F}} + {}^{\mathbf{B}_{3j}} \mathbf{U}_{\mathbf{B}_{\mathrm{4j}}} {}^{\mathbf{B}_{3j}} \boldsymbol{F}\\
    {}^{\mathbf{B}_{4 \mathrm{j}} } \boldsymbol{F}&=
    {}^{\mathbf{B}_{4j}} \hat{\boldsymbol{F}} + {}^{\mathbf{B}_{4j}} \mathbf{U}_{\mathbf{O}_{\mathrm{0}}} {}^{\mathbf{P}_{1j}} \boldsymbol{F}\\
\end{aligned}
\right.
\label{eq:Force_in_frames}
\end{equation}

Also, the net force exerted on the driven point of the revolute segment can be calculated as:
\begin{equation}
\begin{aligned}    
{}^{\mathbf{B}_{\mathbf{cj}}} \boldsymbol{F} = &{}^{\mathbf{B}_{\mathbf{0j}}} \hat{\boldsymbol{F}}
+ { }^{\mathbf{B}_{0 \mathbf{j}}} \mathbf{U}_{\mathbf{B}_{\mathbf{1j}}} {}^{\mathbf{B}_{\mathbf{1j}}} \hat{\boldsymbol{F}}
+ { }^{\mathbf{B}_{2 \mathbf{j}}} \mathbf{U}_{\mathbf{B}_{\mathbf{3j}}} {}^{\mathbf{B}_{\mathbf{3j}}} \hat{\boldsymbol{F}} \\
&+{ }^{\mathbf{B}_{2 \mathbf{j}}} \mathbf{U}_{\mathbf{B}_{\mathbf{3j}}} 
{ }^{\mathbf{B}_{3 \mathbf{j}}} \mathbf{U}_{\mathbf{B}_{\mathbf{4j}}} {}^{\mathbf{B}_{\mathbf{4j}}} \hat{\boldsymbol{F}} \\
&+{ }^{\mathbf{B}_{0 \mathbf{j}}} \mathbf{U}_{\mathbf{B}_{\mathbf{1j}}} 
{ }^{\mathbf{B}_{1 \mathbf{j}}} \mathbf{U}_{\mathbf{B}_{\mathbf{c3}}} {}^{\mathbf{B}_{\mathbf{c3}}} \hat{\boldsymbol{F}}
\end{aligned}
\end{equation}

Given the force ${}^{\mathbf{B}_{c \mathrm{3}} } \boldsymbol{F}$ on the revolute segment, previously calculated for other subsystems in \eqref{eq:Force_in_frames}, the linear actuator force can be determined as follows:
\begin{equation}
\begin{aligned}
f_{c j}=&\mathbf{x}_f^T \, {}^{\mathbf{B}_{4 \mathbf{j}}} \hat{\boldsymbol{F}}-\frac{\mathbf{y}_f^T\left({}^{\mathbf{B}_{4 \mathbf{j}}} \hat{\boldsymbol{F}}\right)\left(x_j+x_{j 0}-l_{c j}\right)}{\left(x_j+x_{j 0}\right) \tan q_{j 2}}\\
&-\frac{\mathbf{z}_\tau^T\left({ }^{\mathbf{B}_{3 \mathbf{j}}} \hat{\boldsymbol{F}}\right)+\mathbf{z}_\tau^T\left({}^{\mathbf{B}_{\mathbf{4 j}}} \hat{\boldsymbol{F}}\right)}{\left(x_j+x_{j 0}\right) \tan q_{j 2}} \\
&-\frac{\mathbf{z}_\tau^T\left({}^{\mathbf{B}_{1 \mathbf{j}}} \hat{\boldsymbol{F}}+{ }^{\mathbf{B}_{1 \mathbf{j}}} \mathbf{U}_{\mathbf{B}_{\mathbf{c3}}}   {}^{\mathbf{B}_{\mathbf{c} \mathbf{3}}} \boldsymbol{F}\right)}{L_{j 1} \sin q_{j 2}}
\end{aligned}
\end{equation}

With ${}^{\mathbf{B}_{\mathbf{c 1}}} \boldsymbol{F}$ now available, the dynamics of object 0 can be derived as follows:
\begin{equation}
\left\{
\begin{aligned}
    {}^{\mathbf{O}_{0}} \boldsymbol{F}&=
    {}^{\mathbf{O}_{0}} \hat{\boldsymbol{F}} +
    {}^{\mathbf{O}_{0}} \mathbf{U}_{\mathbf{B}_{\mathrm{c 1}}} {}^{\mathbf{B}_\mathrm{c 1}} \boldsymbol{F} \\
    {}^{\mathbf{G}} \boldsymbol{F}&={}^{\mathbf{G}} \mathbf{U}_{\mathbf{O}_{ \mathrm{0}}} {}^{\mathbf{O}_{0}} \boldsymbol{F}
\end{aligned}
\right.
\end{equation}
%
\section{Trajectory Optimization for Dynamic Heavy-Duty Manipulator Motion}
\label{section:trajectory_optimization}
This paper focuses on the application of EMLAs as the driving force for parallel-serial manipulators in heavy-duty industrial settings. Each module can execute either a prismatic motion, resembling that of a telescope, or a motion via a parallel mechanism comprising a closed kinematic chain consisting of three rotational-passive joints and a prismatic-actuated joint. We employ a direct collocation strategy using B-spline curves to generate optimal trajectories within our approach. B-splines represent a collection of piecewise polynomial functions constructed iteratively using a knot vector and a set of control points. In this section, we outline the transcription procedure used to transform the optimal control problem into a finite-dimensional nonlinear programming problem (NLP).

To discretize the time variable, we adopt the method described in (\ref{time}), which involves utilizing these time points as collocation points for our problem:
\newcommand{\deleq}{\mathrel{\stackrel{\mathrm{\Delta}}{=}}}
\begin{equation}
\mathcal{T} \deleq \left\{t_0 \cdots t_k \cdots t_M\right\}
\label{time} 
\end{equation}
where $t$ represents the collocation points and $M$ partitions. The configuration vector $\boldsymbol{q} \in \mathbb{R}^{n_a}$, along with its first two time derivatives $\dot{\boldsymbol{q}}$ and $\ddot{\boldsymbol{q}}$, are parameterized using B-spline curves, as defined in (\ref{equation:B_Spline}). In this context, $n_a$ represents the number of DoFs, which refers to the count of actuated joints.

\begin{equation}
\boldsymbol{q} = \boldsymbol{B}(t)\boldsymbol{c}, \quad \dot{\boldsymbol{q}} = \dot{\boldsymbol{B}}(t)\boldsymbol{c}, \text{ and} \quad \ddot{\boldsymbol{q}} = \ddot{\boldsymbol{B}}(t)\boldsymbol{c}
\label{equation:B_Spline} 
\end{equation}

It is noteworthy that passive joints can be expressed as functions of actuated joints. The control points are denoted as $\boldsymbol{c}$, and the basis functions of the B-spline is denoted as $\boldsymbol{B}(t)$ \cite{paz2019practical}. The first and second time derivatives of this basis are thus expressed as $\dot{\boldsymbol{B}}(t)$ and $\ddot{\boldsymbol{B}}(t)$, respectively. Employing the inverse dynamics-based method outlined in \cite{petrovic2022mathematical}, we can derive the velocities $\boldsymbol{v}_{x}$ and forces $\boldsymbol{f}_{x}$ in the linear actuators. This is accomplished through a RNEA designed for handling closed kinematic chains {(\ref{equation:rnea})}.
\begin{equation}
[\boldsymbol{v}_{\!x},\boldsymbol{f}_{\!x}] \ = \ \boldsymbol{R\!N\!E\!A}(\boldsymbol{q},\dot{\boldsymbol{q}},\ddot{\boldsymbol{q}})
\label{equation:rnea}
\end{equation}

Therefore, the NLP problem that must be addressed can be characterized by equations (\ref{equation:costf2}) and (\ref{equation:const2}).
\begin{eqnarray}
\hspace*{-0.92cm}\bld{c}^{*}(\bld{\omega}) \, \leftarrow \, \underset{\boldsymbol{c}, \ t_M}{\operatorname{arg \ min}} & \hspace*{-1.2cm} & f(\boldsymbol{w},\boldsymbol{c}) \ = \ \boldsymbol{w}^{\top} \boldsymbol{\psi}(\boldsymbol{c})
\label{equation:costf2} \\
\nonumber \\
\hspace*{-0.92cm}\mbox{subject to} & \hspace*{-1.2cm} &
\left\{\begin{array}{lcl}
\boldsymbol{q}_{L} & \hspace*{-1.6cm} \leq & \hspace*{-0.8cm} \boldsymbol{q}(t,\boldsymbol{c}) \ \leq \ \boldsymbol{q}_{U} \\
\dot{\boldsymbol{q}}_{L} & \hspace*{-1.6cm} \leq & \hspace*{-0.8cm} \dot{\boldsymbol{q}}(t,\boldsymbol{c}) \ \leq \ \dot{\boldsymbol{q}}_{U} \\
\boldsymbol{f}_{\!xL} & \hspace*{-1.6cm} \leq & \hspace*{-0.8cm} \boldsymbol{f}_{\!x}(t,\boldsymbol{c}) \leq \ \boldsymbol{f}_{\!xU} \\
\boldsymbol{v}_{\!xL} & \hspace*{-1.6cm} \leq & \hspace*{-0.8cm} \boldsymbol{v}_{\!x}(t,\boldsymbol{c}) \leq \ \boldsymbol{v}_{\!xU} \\
t_{M_{L}} & \hspace*{-1.6cm} \leq & \hspace*{-0.4cm} t_M \hspace*{0.24cm} \leq \ t_{M_{U}} \\
\boldsymbol{q}(t_0,\boldsymbol{c}) & \!\!\! = & \!\!\! \boldsymbol{q}_I \\
\boldsymbol{q}(t_M,\boldsymbol{c}) & \!\!\! = & \!\!\! \boldsymbol{q}_F \\
\dot{\boldsymbol{q}}(t_0,\boldsymbol{c}) & \!\!\! = & \!\!\! \dot{\boldsymbol{q}}_I \\
\dot{\boldsymbol{q}}(t_M,\boldsymbol{c}) & \!\!\! = & \!\!\! \dot{\boldsymbol{q}}_F
\end{array} \right.
\label{equation:const2}
\end{eqnarray}

In this case, $\boldsymbol{q}_{L}$, $\dot{\boldsymbol{q}}_{L}$, $\boldsymbol{f}_{x_{L}}$, $\boldsymbol{v}_{x_{L}}$, and $t_{M_{L}}\geq0$ represent the lower limits of joint positions, joint velocities, actuator forces, actuator velocities, and final time, respectively. Conversely, $\boldsymbol{q}_{U}$, $\dot{\boldsymbol{q}}_{U}$, $\boldsymbol{f}_{x_{U}}$, $\boldsymbol{v}_{x_{U}}$, and $t_{M_{U}}$ represent their corresponding upper limits. The initial and final configurations, $\boldsymbol{q}_I$ and $\boldsymbol{q}_F$, along with their first time derivatives, $\dot{\boldsymbol{q}}_I$ and $\dot{\boldsymbol{q}}_F$, are enforced through the last four constraints in \eqref{equation:const2}. Clearly, if we set $t_{0}=0$, then the total time required to complete the task is $t_{M}$. Furthermore, the cost function in (\ref{equation:costf2}) is formulated as the dot product between a weight vector $\boldsymbol{w} \in \mathbb{R}^e$ and its corresponding variables, with the condition as \eqref{eq: weighting vector}:
\begin{equation}
\boldsymbol{w} \ = \ \begin{bmatrix} w_1 & w_2 & ... & w_e \end{bmatrix}^\top
\label{eq: weighting vector}
\end{equation}
and a vector of criteria $\boldsymbol{\psi}(t,\boldsymbol{c}) \in \mathbb{R}^e$, defined as (\ref{equation:criteria}):
\begin{equation}
\boldsymbol{\psi}(t,\boldsymbol{c}) \ = \ \begin{bmatrix} \psi_1(t,\boldsymbol{c}) & \psi_2(t,\boldsymbol{c}) &
... & 
\psi_e(t,\boldsymbol{c})
\end{bmatrix}^\top
\label{equation:criteria}
\end{equation}

As the case study, selected EMLAs of Table \ref{EMLA_Table}, previously presented in Section \ref{section:energy_conversion}, are utilized to actuate the 3-DoF mechanism reported in \cite{petrovic2022mathematical}. Regarding manipulator optimization, two scalar criteria $\psi_1(\cdot)$ and $\psi_2(\cdot)$ are combined linearly with the scalar weights $w_1$ and $w_2$ to enable a multi-objective problem, and the specific definitions of these criteria are as follows.

\begin{itemize}
\item \textit{Minimum effort in actuators:} The objective is to optimize the forces of the linear actuators, achieved by minimizing the quadratic form of forces described in (\ref{equation:minimumeffort}):
\begin{equation}
\psi_{\text{1}} \ = \ \tfrac{1}{2}  \Delta_t\sum_{t=t_0}^{t_M} \boldsymbol{f}_{\!x}^{\top} \boldsymbol{f}_{\!x} 
\label{equation:minimumeffort}
\end{equation}
where $\Delta_t = t_k-t_{k-1}$ and $\boldsymbol{f}_{\!x}$ is computed through (\ref{equation:rnea}).
\item \textit{Minimum required mechanical power:} The goal is to minimize the output power provided by the EMLA over the whole trajectory, as obtained in {(\ref{equation:minimumpower})}: 
\begin{equation}
\psi_{\text{2}} \ = \ \tfrac{1}{2} \Delta_t \sum_{t=t_0}^{t_M} \sum_{i=1}^{n_{a}} \left( f_{\!x_{i}} v_{x_{i}} \right)^{2}
\label{equation:minimumpower}
\end{equation}

Within this context, $p_i = f_{\!x_{i}} v_{x_{i}} \in\mathbb{R}$ embodies the mechanical output power associated with the implemented actuator $i$ at time $t_k$. Here, $f_{x_{i}}$ and $v_{x_{i}}$ denote the $i$-th component of $\boldsymbol{f}_{\!x}$ and $\boldsymbol{v}_{\!x}$, respectively, which are retrieved from (\ref{equation:rnea}).
\end{itemize}
\section{Bilevel Multi-Objective Optimization Concept}
\label{section:bilevel_optimization}
Multi-objective optimization tackles the simultaneous optimization of conflicting objectives. In contrast, multi-level optimization addresses problems with interconnected decision-making, where follower's decisions impact those at the leader level \cite{dempe2020bilevel}. By merging these frameworks, the concept of multi-level multi-objective optimization emerges as a powerful decision-making tool for complex systems characterized by multiple competing objectives \cite{10121701,eichfelder2010multiobjective,beck2023survey,fliege2006multicriteria,dempe2020bilevel}. This approach offers valuable insights for decision support, enabling the identification of balanced solutions in practical applications \cite{10459711,barahimi2021multi,bard2013practical}.
\subsection{Problem Formulation with a Leader--Follower Scenario}
\label{multi_objective_optimization_problem_formulation}
This section delves into the formulations for a bilevel multi-objective optimization problem, encompassing two distinct decision-making levels. The outer level represents a leader's perspective, while the inner level reflects that of a follower. We specifically practice a formulation where the decision maker, operating at the inner level, exerts influence by selecting a solution that aligns with their own preferences, effectively shaping the feasible set of solutions at the outer level. Accordingly, there are two types of variables in this problem, namely, the outer level variables $\bld{x}_{\!_{O}}$ and the inner level variables $\bld{x}_{\!_{I}}$.
\\
{\bf{Definition.}} The constrained multi-objective inner optimization problem is expressed by \eqref{equation:costf2_i}-\eqref{equation:const2_i}:
\begin{eqnarray}
\hspace*{-0.92cm}\bld{\xi}^{*}(\bld{x}_{\!_{O}}) \, \leftarrow \, \underset{\bld{x}_{\!_{I}}}{\operatorname{arg \ min}} & \hspace*{-1.2cm} & f(\bld{x}_{\!_{O}},\bld{x}_{\!_{I}}) \ = \ \sum_{i=1}^{e} f_{i}(\bld{x}_{\!_{O}},\bld{x}_{\!_{I}})
\label{equation:costf2_i} \\
\nonumber \\
\hspace*{-0.92cm}\mbox{subject to} & \hspace*{-1.2cm} & \bld{g}(\bld{x}_{\!_{O}},\bld{x}_{\!_{I}}) \ \leq \ \bld{0}
\label{equation:const2_i}
\end{eqnarray}
where $f(\cdot)$ is the cost function, composed of the sum of criteria $f_{i}(\cdot)$, and $\bld{g}(\cdot)$ contains all unilateral and bilateral constraints. Therefore, the optimal argument $\bld{\xi}^{*}$ is the shared variable that connects the inner with the outer level. This inner level can be defined as an optimal control problem, transcribed into an NLP similar to that defined in (\ref{equation:costf2})-(\ref{equation:const2}), which results in an optimal trajectory generator for the robotic manipulator.

\begin{figure}[btp]
    \centering
    \includegraphics[trim={0.0cm 0.0cm 0.0cm 0.0cm},clip,width=9.25cm]{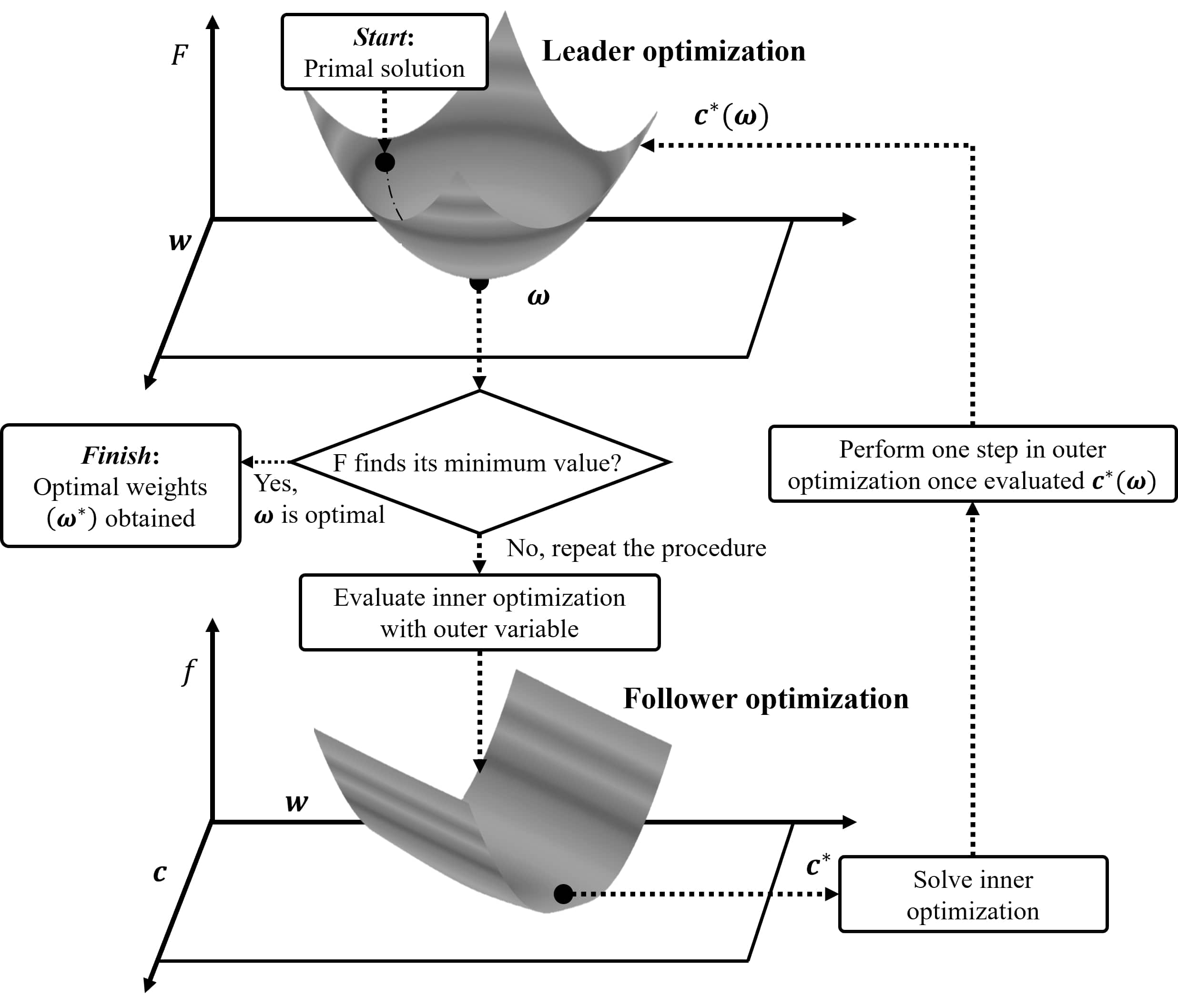}
    \caption{Flowchart of the bilevel optimization through leader--follower approach}
    \label{fig:Bilevel_Flowchart}
\end{figure}

The outer level is thus defined by the following constrained optimization problem, as \eqref{equation:costf2_o}-\eqref{equation:const2_o}:
\begin{eqnarray}
\hspace*{-0.92cm}\bld{x}_{\!_{O}}^{*} \, \leftarrow \, \underset{\bld{x}_{\!_{O}}, \ \bld{x}_{\!_{I}}}{\operatorname{minimize}} & \hspace*{-1.2cm} & \hspace*{-4.0cm} F(\bld{x}_{\!_{O}},\bld{x}_{\!_{I}})
\label{equation:costf2_o} \\
\nonumber \\
\hspace*{1.2cm}\mbox{subject to} & 
\left\{\begin{array}{lcl}
\hspace*{1.15cm} \bld{x}_{\!_{I}} & \in & \bld{\xi}(\bld{x}_{\!_{O}}) \\
\bld{G}(\bld{x}_{\!_{O}},\bld{x}_{\!_{I}}) & \leq & \bld{0}
\end{array} \right.
\label{equation:const2_o}
\end{eqnarray}
where $F(\cdot)$ is the cost function to optimize subject to the general constraints $\bld{G}(\cdot)$ and forcing $\bld{x}_{\!_{I}}$ belonging to the set of optimal solutions of the inner level (\ref{equation:costf2_i})-(\ref{equation:const2_i}). The interaction between the leader level and follower level in bilevel optimization problems is illustrated in Fig. \ref{fig:Bilevel_Flowchart}.
Motivated by the need to balance the dynamic interplay and address the multi-disciplinary problem mentioned in Section \ref{Motivations}, we adopt the bilevel multi-objective optimization algorithm (\ref{equation:costf2_o})-(\ref{equation:const2_o}) to simultaneously optimize across multiple objectives associated with the manipulator and EMLA performance in Section \ref{case_study}.
\subsection{EMLA-Driven Robotic Manipulator Optimization}
\label{case_study}
\subsubsection{Algorithm Structure}
This section proposes a holistic optimization paradigm for an electrified $n_a$-DoF robotic manipulator with EMLAs integrated within its joints. The approach formulates the problem as a leader--follower bilevel multi-objective optimization framework, building upon the groundwork laid in Section \ref{multi_objective_optimization_problem_formulation}. 
To illustrate the developed bilevel optimizer, \textbf{Algorithm 1} presents its structure. The inner level optimizer operates from lines 15 to 36, while lines 1 to 14 represent the outer-level optimizer. The inner level, which is set out in \eqref{equation:costf2_i}-\eqref{equation:const2_i}, features the manipulator's trajectory generator, which iteratively refines its decisions to generate optimal trajectories in the configuration space considering the inner level cost function (\textbf{InCost}) and manipulator constraints (\textbf{InConstraints}). Concurrently, the outer level, with its architecture defined in \eqref{equation:costf2_o}-\eqref{equation:const2_o}, prioritizes achieving the highest possible peak actuator power conversion (\textbf{OutCost}) while adhering to their operational constraints (\textbf{OutConstraints}). In this case study, the effort in the pistons of the studied manipulator and the delivered power of the EMLA at the load side is defined as \textbf{InCost} through \eqref{equation:minimumeffort}-\eqref{equation:minimumpower} (detailed in Section \ref{section:trajectory_optimization}) in order to generate the optimal trajectory. On the other side, maximizing the efficiency of the three EMLAs for the highest possible performance is defined as \textbf{OutCost} using \eqref{system_efficiency} (detailed in Section \ref{section:energy_conversion}).
\begin{table}[h]
  \centering
  \begin{threeparttable}
  \renewcommand{\arraystretch}{0.9} 
  \begin{tabular}{p{0.9\linewidth}} 
    \hline
    \\
    \multicolumn{1}{c}{\textbf{Algorithm 1}: Leader-Follower Bilevel Optimizer} \\
    \\
    \hspace{0.4cm}\textbf{Input}: $n_a$, $N$ and $M$ .\\
    \hspace{0.4cm}\textbf{Output}: Optimal weights $\bld{\omega}^{*}$.\\
    \\
    {\small 1}\hspace{0.2cm}$\bld{\omega}^{*} \leftarrow$ \textbf{OuterOptimizer}($n_a$, $N$, $M$, \textbf{OutCost}, \textbf{OutConstraints})\\
    {\small 2}\hspace{0.2cm}\textbf{OutCost}($n_a$, $N$, $M$, $\bld{\omega}$, $t_M$)\\
    {\small 3}\hspace{0.6cm}$\bld{c}^{*}, t_M \leftarrow$ \textbf{InnerOptimizer}($n_a, N, M, \bld{\omega}$,\textbf{InCost},\textbf{InConstraints})\\
    {\small 4}\hspace{0.6cm}$F \leftarrow 0$\\
    {\small 5}\hspace{0.6cm}$\boldsymbol{B}$, $\dot{\boldsymbol{B}}$, $\ddot{\boldsymbol{B}}\leftarrow$ \textbf{BasisFunctions}($n_a$, $N$, $M$, $t_M$)\\
    {\small 6}\hspace{0.6cm}\textbf{For} $k=0:M$ \textbf{then}\\
    {\small 7}\hspace{1.0cm}$\bld{q}=\bld{B}(t_k)\bld{c}^{*}$ \quad $\dot{\bld{q}}=\dot{\bld{B}}(t_k)\bld{c}^{*}$ \quad $\ddot{\bld{q}}=\ddot{\bld{B}}(t_k)\bld{c}^{*}$\\
    {\small 8}\hspace{1.0cm}$[\bld{v}_{\!x}$, $\bld{f}_{\!x}] \leftarrow$ \textbf{RNEA}($\bld{q}$, $\dot{\bld{q}}$, $\ddot{\bld{q}}$)\\
    {\small 9}\hspace{1.0cm}$F \leftarrow F - \tfrac{1}{2}  \Delta_t \left(\bld{\eta}_{\textbf{Act}} (\bld{v}_{\!x}, \bld{f}_{\!x})\right)^2$\\
    {\small 10}\hspace{0.4cm} \textbf{end}\\
    {\small 11}\hspace{0.0cm} \textbf{return} $F$\\
    {\small 12}\hspace{0.1cm}\textbf{OutConstraints}($n_a$, $N$, $M$, $\bld{\omega}$, $t_M$)\\
    {\small 13}\hspace{0.5cm}Set $\bld{\omega}_{\!L} \leq \bld{\omega} \leq \bld{\omega}_{\!U}$ \\
    {\small 14}\hspace{0.0cm} \textbf{return} Stack of constraints\\
    {\small 15}\hspace{0.1cm}\textbf{InCost}($n_a$, $N$, $M$, $\bld{\omega}$, $\bld{c}$, $t_M$)\\
    {\small 16}\hspace{0.5cm}$f \leftarrow 0$\\
    {\small 17}\hspace{0.5cm}$\boldsymbol{B}$, $\dot{\boldsymbol{B}}$, $\ddot{\boldsymbol{B}}\leftarrow$ \textbf{BasisFunctions}($n_a$, $N$, $M$, $t_M$)\\
    {\small 18}\hspace{0.5cm}\textbf{For} $k=0:M$ \textbf{then}\\
    {\small 19}\hspace{0.9cm}$\bld{q}=\bld{B}(t_k)\bld{c}$ \quad $\dot{\bld{q}}=\dot{\bld{B}}(t_k)\bld{c}$ \quad $\ddot{\bld{q}}=\ddot{\bld{B}}(t_k)\bld{c}$\\
    {\small 20}\hspace{0.9cm}$[\bld{v}_{\!x}$, $\bld{f}_{\!x}] \leftarrow$ \textbf{RNEA}($\bld{q}$, $\dot{\bld{q}}$, $\ddot{\bld{q}}$)\\
    {\small 21}\hspace{0.9cm}Compute criteria $\psi_{1}(\bld{c})$ and $\psi_{2}(\bld{c})$\\
    {\small 22}\hspace{0.9cm}$f \leftarrow f + \bld{\omega}^{\top} \bld{\Psi}$\\
    {\small 23}\hspace{0.5cm} \textbf{end}\\
    {\small 24}\hspace{0.0cm} \textbf{return} $f$\\
    {\small 25}\hspace{0.1cm}\textbf{InConstraints}($n_a$, $N$, $M$, $\bld{c}$, $t_M$)\\
    {\small 26}\hspace{0.5cm}$\bld{B}$, $\dot{\bld{B}}$, $\ddot{\bld{B}}\leftarrow$ \textbf{BasisFunctions}($n_a$, $N$, $M$, $t_M$)\\
    {\small 27}\hspace{0.5cm}Set $\bld{B}(t_0)\bld{c} = \bld{q}_{I}$ and $\bld{B}(t_M)\bld{c} = \bld{q}_{F}$\\
    {\small 28}\hspace{0.5cm}Set $\dot{\bld{B}}(t_0)\bld{c} = \dot{\bld{q}}_{I}$ and $\dot{\bld{B}}(t_M)\bld{c} = \dot{\bld{q}}_{F}$\\
    {\small 29}\hspace{0.5cm}\textbf{For} $k=0:M$ \textbf{then}\\
    {\small 30}\hspace{0.9cm}$\bld{q}=\bld{B}(t_k)\bld{c}$ \quad $\dot{\bld{q}}=\dot{\bld{B}}(t_k)\bld{c}$ \quad $\ddot{\bld{q}}=\ddot{\bld{B}}(t_k)\bld{c}$\\
    {\small 31}\hspace{0.9cm}$[\bld{v}_{\!x}$, $\bld{f}_{\!x}] \leftarrow$ \textbf{RNEA}($\bld{q}$, $\dot{\bld{q}}$, $\ddot{\bld{q}}$)\\
    {\small 32}\hspace{0.9cm}Set $\boldsymbol{q}_{L} \leq \boldsymbol{q} \leq \boldsymbol{q}_{U}$ and $\dot{\boldsymbol{q}}_{L} \leq \dot{\boldsymbol{q}} \leq \dot{\boldsymbol{q}}_{U}$\\
    {\small 33}\hspace{0.9cm}Set $\bld{v}_{\!x_{L}} \leq \bld{v}_{\!x} \leq \bld{v}_{\!x_{U}}$ and $\bld{f}_{\!x_{L}} \leq \bld{f}_{\!x} \leq \bld{f}_{\!x_{U}}$\\
    {\small 34}\hspace{0.9cm}Set $t_{\!M_{L}} \leq t_{M} \leq t_{\!M_{U}}$\\
    {\small 35}\hspace{0.5cm} \textbf{end}\\
    {\small 36}\hspace{0.0cm} \textbf{return} Stack of constraints\\
    \\
    \hline
  \end{tabular}
    \begin{tablenotes}  
      \item[- Superscript asterisk (*) indicates optimal values obtained within the optimization framework.] 
      \item[- Bold symbols in uppercase/lowercase represent matrices/vectors. The rest denote scalar quantities.] 
    \end{tablenotes}
  \end{threeparttable}
\end{table}
\subsubsection{Optimization Methodology}
At the outset, the optimal trajectory generator, defined in \eqref{equation:costf2}-\eqref{equation:const2} of Section \ref{section:trajectory_optimization}, provides optimal control points as a function of the weighting factors, i.e. $\bld{c}^{*}(\bld{\omega})$. In order to find optimal values for such weights, the trajectory generator can be embedded into a bilevel framework as the inner level optimizer. Therefore, in the outer level optimizer, an EMLAs efficiency function can be maximized subject to constraints on the weights. Following the aforementioned description, the bilevel multi-objective optimizer for our problem is defined by (\ref{eq:costf2__})-(\ref{eq:const2__}) where the control points of the B-Spline $\bld{c}$ is the shared variable that connects the inner level with the outer level optimizer.
	\begin{eqnarray}
		\hspace*{-0.8cm}\underset{\bld{\omega}}{\operatorname{maximize}} & \ & \hspace*{-0.2cm}F(\bld{\omega}) \ = \ \tfrac{1}{2}  \Delta_t\sum_{t=t_0}^{t_M} \left(\bld{\eta}_{\textbf{Act}} (\bld{v}_{\!x}, \bld{f}_{\!x})\right)^2
		\label{eq:costf2__} \\
		\hspace*{-0.8cm}\nonumber \\
		\hspace*{-0.8cm}\nonumber \\
		\hspace*{-0.8cm}\mbox{subject to} & \ &
		\hspace*{-0.2cm}\left\{\begin{array}{lcl}
			\bld{c} & \in & \bld{c}^{*}(\bld{\omega}) \ \text{from} \ (\ref{equation:costf2})\!\!-\!\!(\ref{equation:const2}) \\
			\bld{\omega}_{L} & \leq & \bld{\omega} \ \ \leq \ \ \bld{\omega}_{U} \\
		\end{array} \right.
		\label{eq:const2__}
	\end{eqnarray}
where $\bld{\omega}$ is the weights vector, with boundaries $\bld{\omega}_{L}$ and $\bld{\omega}_{U}$, and $F(\bld{\omega})$ stands for the efficiency over time of all EMLAs installed in the manipulator's joints. At instant $t_k$, the EMLAs' efficiency is computed through $\bld{\eta}_{\textbf{Act}} (\bld{v}_{\!x}, \bld{f}_{\!x}):\mathbb{R}^{2n_a}\rightarrow\mathbb{R}$, which is a scalar function defined as \eqref{outer_obj}.
\begin{equation}
    \bld{\eta}_{\textbf{Act}} (\bld{v}_{\!x}, \bld{f}_{\!x}) \ = \ \frac{\bld{v}_{\!x}^{\top} \bld{f}_{\!x}}{\sum_{i=1}^{n_{a}} \frac{f_{\!x_{i}} v_{x_{i}}} {\eta_{\text{EMLA}} ( f_{\!x_{i}}, v_{x_{i}} )}} 
\label{outer_obj}
\end{equation}
where the single-EMLA efficiency is retrieved with $\eta_{\text{EMLA}}( f_{\!x}, v_{x} ):\mathbb{R}^{2}\rightarrow\mathbb{R}$ defined in (\ref{system_efficiency}). Please note that, since $\bld{v}_{\!x}$ and $\bld{f}_{\!x}$ are functions of $\bld{c}$, which is a function of $\bld{\omega}$, then $F$ is a function of $\bld{\omega}$. Also, the first constraint forces the control points for belonging to the set of optimal solutions of the trajectory generator.
\subsubsection{Results and Discussions}
\label{results_and_discussions}
The optimization algorithm developed in Section \ref{case_study} offers distinct advantages that address the challenges described in Section \ref{Motivations}, as below:
\begin{itemize}
    \item In bilevel optimization, which leverages a two-stage approach through \eqref{equation:costf2_i}-\eqref{equation:const2_o}, the follower's decision-making, based on its own objectives and constraints, is guided by the leader's goal and weight limits. This leads to a more streamlined and effective optimization process and a highly accurate solution for such multidisciplinary problems. 
    \item A key strength of the bilevel framework lies in its inherent capability to effectively balance potentially conflicting objectives. Single-level optimizations often struggle where multiple objectives need to be simultaneously optimized. By enabling the outer level to determine optimal weighting factors through \eqref{eq:costf2__}-\eqref{eq:const2__}, it cooperatively harmonizes the objectives, and navigates this challenge.
    \item Furthermore, once the optimal weights are identified, the computational burden is significantly reduced since the inner-level optimizer \eqref{equation:costf2} alone becomes sufficient for generating trajectories that not only meet the desired motion requirements but also guarantee maximum efficiency in the employed EMLAs integrated in the manipulator.
\end{itemize}  

\begin{figure*} [bt]
\centering
\includegraphics[trim={0.0cm 0.0cm 0.0cm 0.0cm},clip,width=\linewidth]{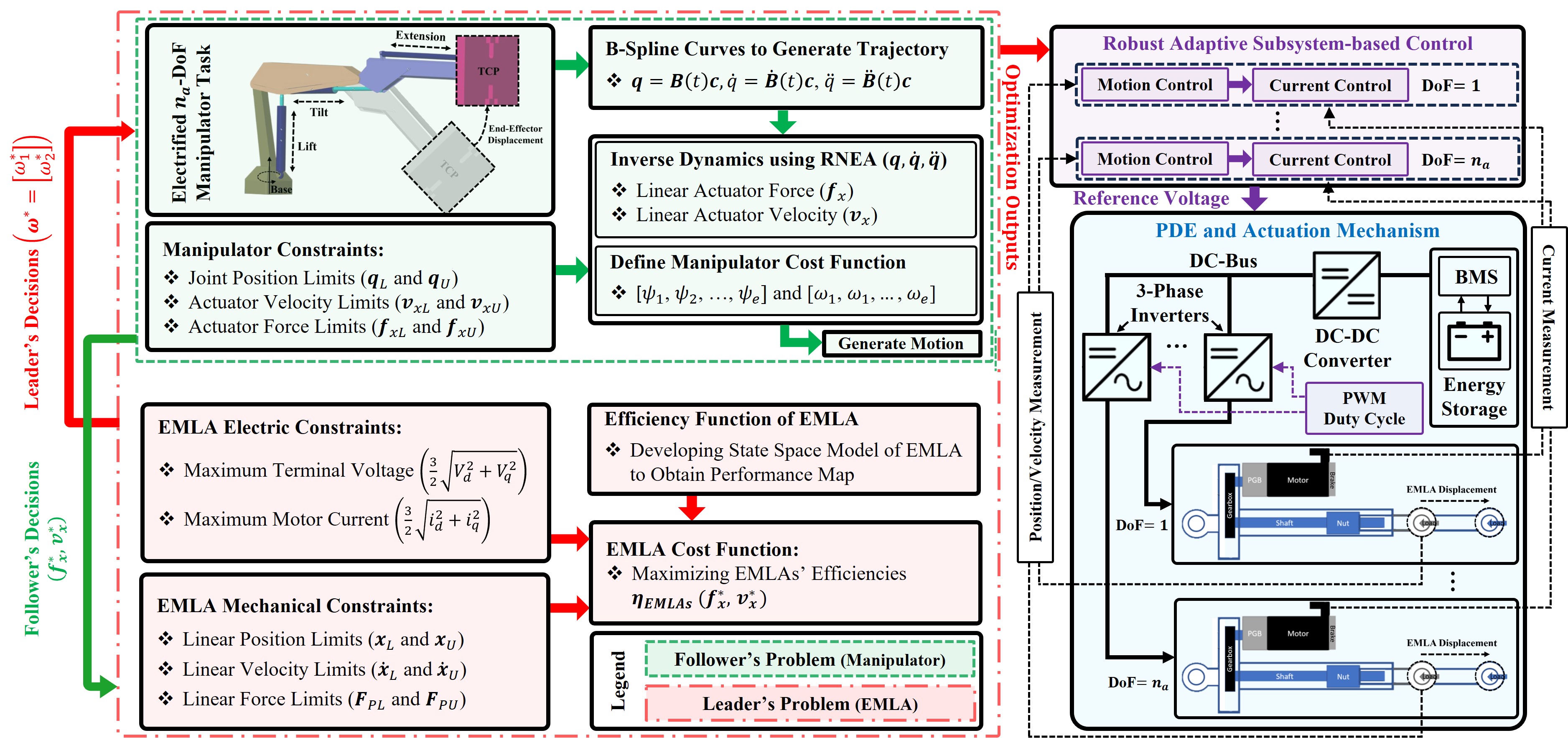}
\caption{Integrated framework for achieving holistically efficient performance of EMLA-driven robotic manipulators through a leader-follower bilevel optimization scenario}
\label{Electrification_Bilevel_Procedure}
\end{figure*}

\begin{figure} [h]
    \centering
    \includegraphics[trim={0.0cm 0.0cm 0.0cm 0.0cm},clip,width=5cm]{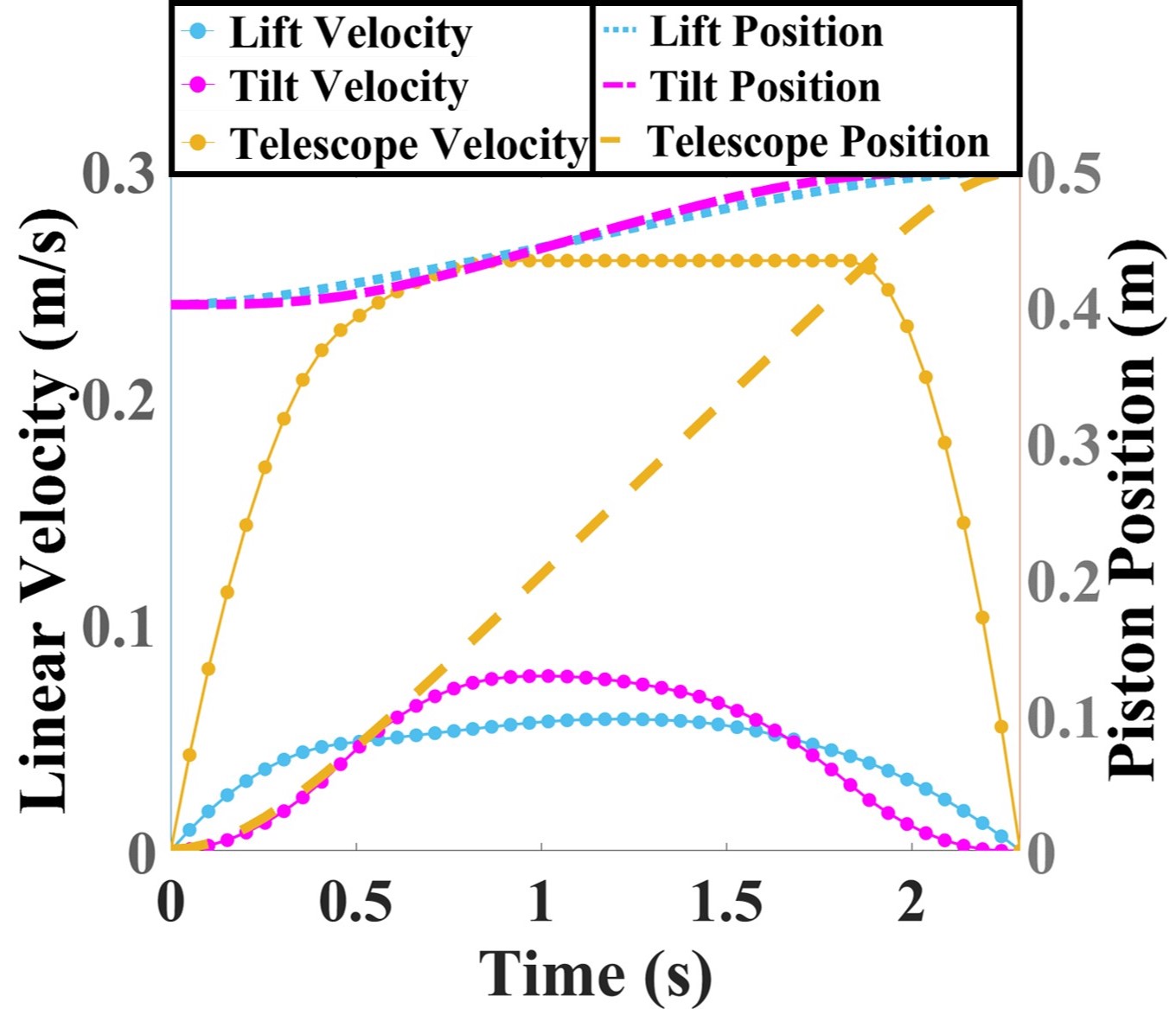}
    \caption{Optimal force and power in the lift, tilt, and telescope joints of the manipulator}
    \label{opt_force_power}
\end{figure}

\begin{figure} [h]
    \centering
    \includegraphics[trim={0.0cm 0.0cm 0.0cm 0.0cm},clip,width=5cm]{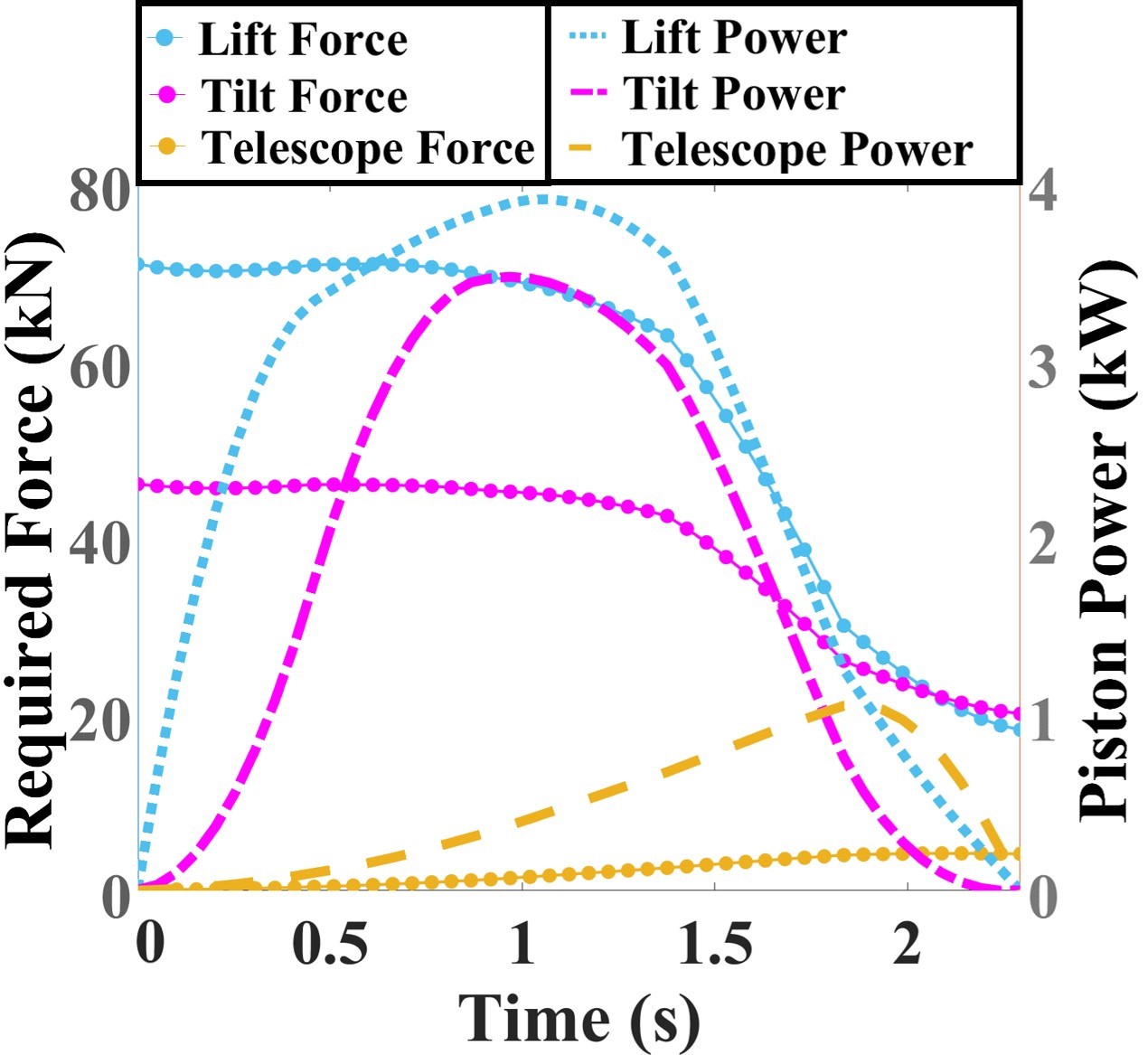}
    \caption{Optimal linear velocity and position in the lift, tilt, and telescope joints of the manipulator}
    \label{opt_velocity_position}
\end{figure}

\begin{figure}[h]
    \centering
    \includegraphics[trim={0.0cm 0.0cm 0.0cm 0.0cm},clip,width=8.5cm]{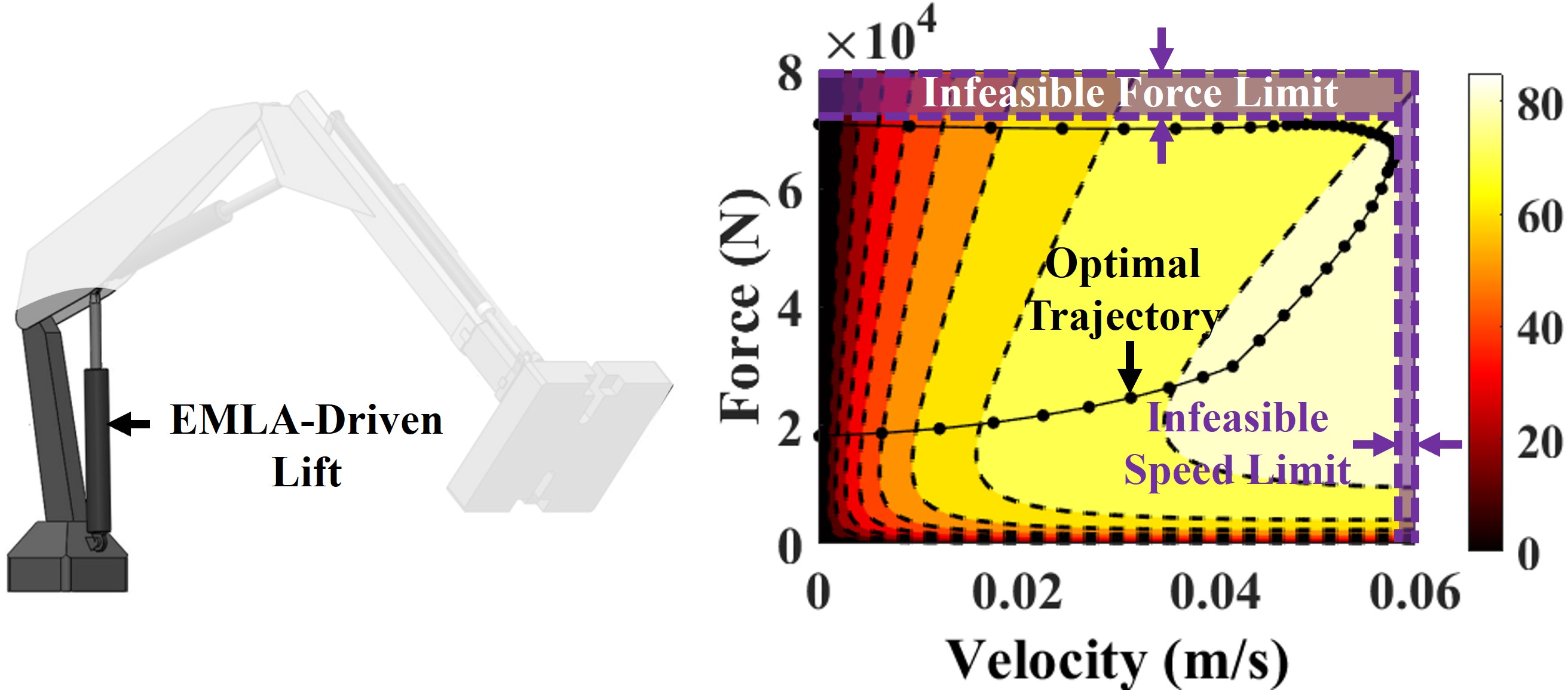}
    \caption{Obtained optimal trajectory of 3-DoF manipulator plotted in efficiency map of EMLA implemented in lift joint}
    \label{Lift_eff}
\end{figure}

\begin{figure}[h]
    \centering
    \includegraphics[trim={0.0cm 0.0cm 0.0cm 0.0cm},clip,width=8.5cm]{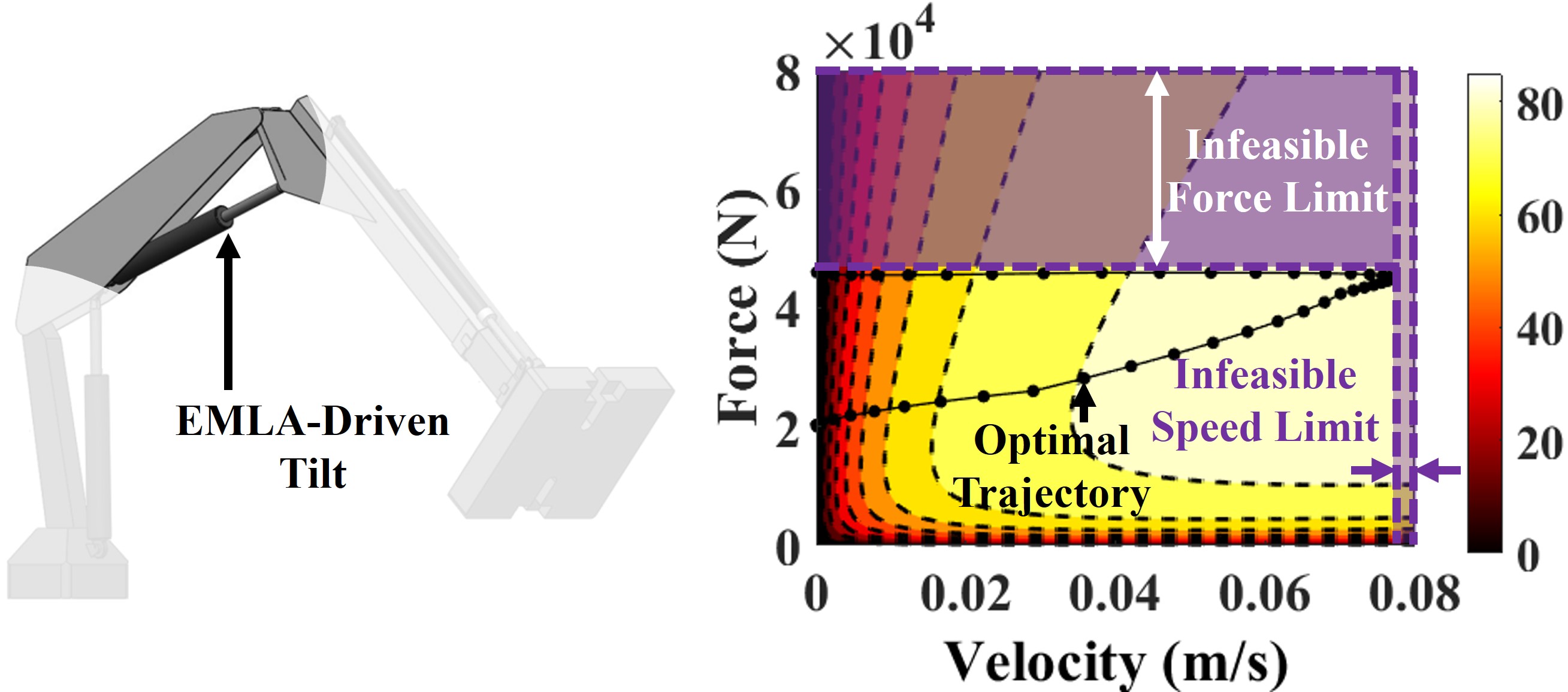}
    \caption{Obtained optimal trajectory of 3-DoF manipulator plotted in efficiency map of EMLA implemented in tilt joint}
    \label{Tilt_eff}
\end{figure}

\begin{figure}[h]
    \centering
    \includegraphics[trim={0.0cm 0.0cm 0.0cm 0.0cm},clip,width=8.5cm]{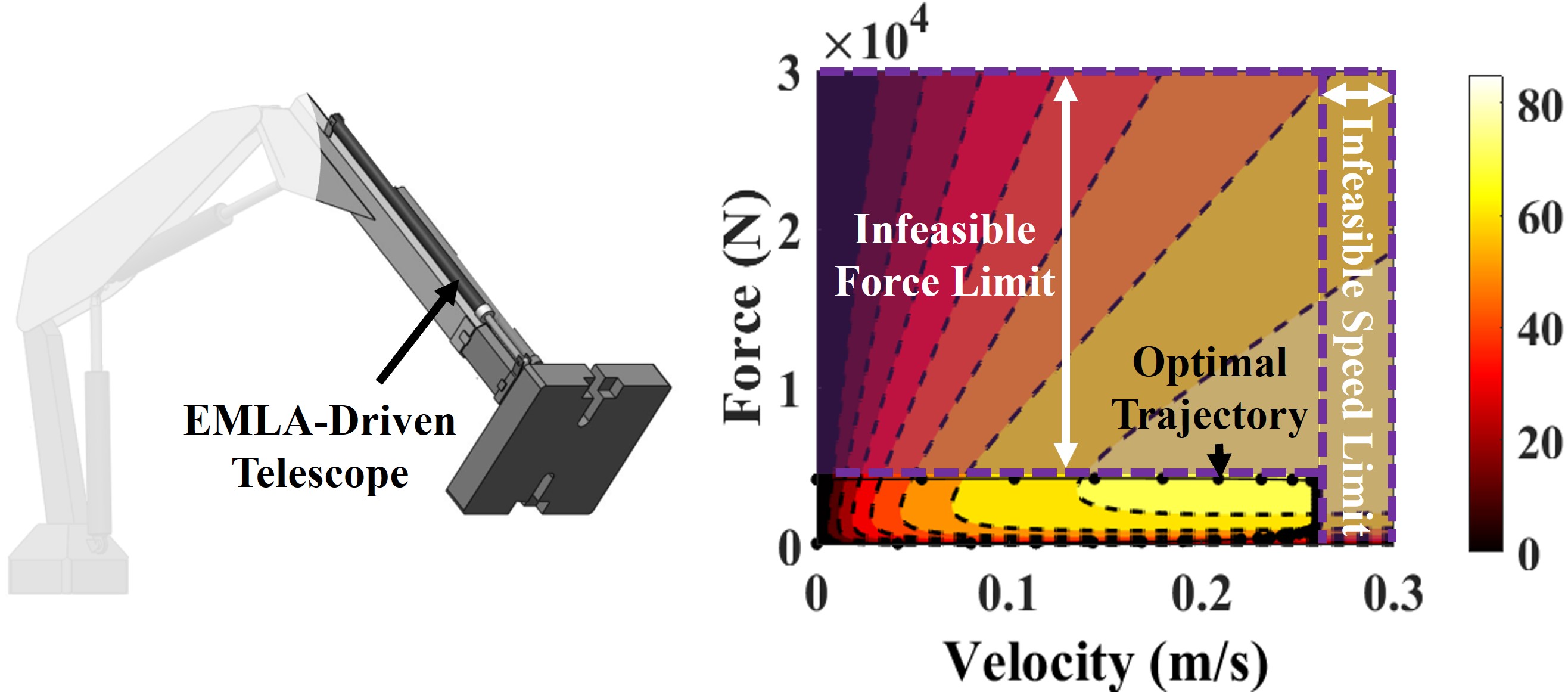}
    \caption{Obtained optimal trajectory of 3-DoF manipulator plotted in efficiency map of EMLA implemented in telescope joint}
    \label{Telescope_eff}
\end{figure}

The adopted bilevel optimization strategy, illustrated in Fig. \ref{Electrification_Bilevel_Procedure}, yields the optimal solution including $\bld{v}_{\!x}$, $\bld{f}_{\!x}$, $\bld{q}$, and $p_i = f_{\!x_{i}}v_{x_{i}}, i\in\{1,\cdots,n_a\}$ of each joint, as shown in Fig. \ref{opt_force_power} and Fig. \ref{opt_velocity_position}. Additionally, Fig. \ref{Lift_eff} to Fig. \ref{Telescope_eff} further complements these results by showcasing the optimal trajectories overlaid on the efficiency maps of the EMLAs within the lift, tilt, and telescope joints. It can observed how the optimized trajectories aligned with the high-efficiency regions of the EMLAs. To quantify, the corresponding efficiencies of the EMLAs are 73.1\%, 66.2\%, and 57.5\%, respectively. Also, the total efficiency of actuation system, defined in \eqref{outer_obj}, achieved 70.3\%.
 
\section{Robust Decomposed System Control}
\label{section:control}
\subsection{Control Strategy for $n_a$-DoF Robotic Manipulator}
To track optimal trajectory achievement and stability in control decision-making, we apply a generic robust decomposed system control (RDSC) approach by developing the strategy proposed in \cite{heydari2024robust} into electrified manipulator dynamics actuated by $n_a$ EMLAs. The primary reason for employing this control algorithm is its ability to achieve uniform exponential stability in convergence rates. In addition, it efficaciously addresses external effects in EMLA systems, such as manufacturing variations, environmental factors, load variations, wear and aging, power supply fluctuations, and sensor inaccuracy \cite{7539392,9802805,8839808}. To tackle unknown disturbances and ensure that steady states track the generated optimal trajectories, we have incorporated boundary estimation and robust terms into the control and parameter adaptive laws, respectively.
Through this approach, we apply two distinct subsystem-based controllers for each EMLA. The first control section is designed to control the linear motion of the EMLAs, with states for the linear position $x_{1,i}$ and the linear velocity $x_{2,i}$ so each $i$th EMLA can follow the optimal trajectory obtained from \textbf{Algorithm 1}. It generates sufficient torque (${\tau_m}$) to ensure the linear velocities of the EMLAs track the optimal linear velocities ($\bld{v}_{\!x}$) outlined in Section \ref{case_study}. In addition, the current control of the motors with states $i_q=x_{3,i}$ and $i_d=x_{4,i}$ for the $i$th motor is elaborated upon in Section \ref{case_study_control}. In this phase, once the required torques are achieved, the reference currents for $x_{{3,i}(ref)}$ are established by considering \eqref{equation:electromagnetictorque} with $x_{{4,i}(ref)}=0$ to achieve the highest torque production. Subsequently, treating currents as states, this control system applies sufficient voltages to ensure that the actual currents align with the required currents. Considering the state space presented in \eqref{state_space_2}, converted into linear motion, we identify four subsystems for each EMLA, resulting in a total of $4 \times n_a$ subsystems. If we consider linear motion and motor current states $[x_{1,i},...,x_{4,i}]^\top$ for the $i$th EMLA that $i=1, 2, ..., n_a$, we introduce a transformation to the tracking form by defining $Q_{\nu,i}$, as follows:
\begin{equation}
\begin{aligned}
\label{equation:39}
Q_{\nu,i} &= \begin{cases} 
x_{\nu,i} - x_{\nu,i_{(ref)}} - \kappa_{\nu-1,i} & \text{if } \nu=2 \\
x_{\nu,i} - x_{\nu,i_{(ref)}} & \text{if } \nu=1,3,4 
\end{cases}
\end{aligned}
\end{equation}
where $\nu=1,\ldots,4$ and $x_{\nu,i}^{\text{(ref)}}$ is the optimal trajectory generated in Section \ref{case_study_control}. To fulfill the aforementioned purposes, we propose the control approach as formulated in \eqref{equation: 41}:
\begin{equation}
\begin{aligned}
\label{equation: 41}
{\kappa_{\nu,i}} = - \frac{1}{2} (\delta_{\nu,i}+\epsilon_{\nu,i}\hat{\phi}_{\nu,i}){Q_{\nu,i}}
\end{aligned}
\end{equation}
where ${\kappa_{1,i}}$ is a virtual control, ${\kappa_{2,i}}={\tau_m}$ is $i$th EMLA torque, ${{\kappa_{3,i}}=V_q}$ and ${{\kappa_{4,i}}=V_d}$ are the $i$th EMLA’s motor voltages in the $q$- and $d$-axis, $\delta_{\nu,i}$ and $\epsilon_{\nu,i}$ are positive constants, and $\hat{\phi}_{\nu,i}$ is an adaptive law provided, as elaborated in \eqref{equation: 45}:
\begin{equation}
\begin{aligned}
\label{equation: 45}
\dot{\hat{\phi}}_{\nu,i} &= -{k_{\nu,i}}{\sigma_{\nu,i}}\hat{\phi}_{\nu,i}+\frac{1}{2}\epsilon_{\nu,i}{k_{\nu,i}}|{Q_{\nu,i}}|^2
\end{aligned}
\end{equation}
where $k_{\nu,i}$ and $\sigma_{\nu,i}$ are positive constants. As observed, the control approach incorporates constant parameters, such as $\delta_{\nu,i}$, $\epsilon_{\nu,i}$, $k_{\nu,i}$, and $\sigma_{\nu,i}$, all of which must be positive.
\begin{table}[h]
\centering
\renewcommand{\arraystretch}{0.9} 
\begin{tabular}{p{0.9\linewidth}} 
\hline
\\
\multicolumn{1}{c}{\textbf{Algorithm 2}: Robust subsystem-based control} \\
\\
\hspace{0.4cm}\textbf{Input}: \textcolor{black}{$x_{\nu,i}$ from sensory sections, $x_{\nu,i}^{(\text{ref})}$ from algorithm 1.}\\
\hspace{0.4cm}\textbf{Output}: Control signals $\kappa_{\nu,i}$.\\
\\
{\small 1}\hspace{0.2cm}\textbf{For} $i=1:n_a$ \textbf{do}\\
{\small 2}\hspace{0.7cm}\textbf{For} $\nu=1:4$ \textbf{do}\\
{\small 3}\hspace{1.2cm} \textbf{If} $\nu = 1$ \textbf{then}\\
{\small 4}\hspace{1.7cm} $Q_{1,i} = x_{1,i} - x_{1,i}^{\text{(ref)}}$;\\
{\small 5}\hspace{1.7cm} $\dot{\hat{\phi}}_{1,i} = -{k_{1,i}}{\sigma_{1,i}}\hat{\phi}_{1,i}+\frac{1}{2}\epsilon_{1,i}{k_{1,i}}|{Q_{1,i}}|^2$;\\
{\small 6}\hspace{1.7cm} ${\kappa_{1,i}} = - \frac{1}{2} (\delta_{1,i}+\epsilon_{1,i}\hat{\phi}_{1,i}){Q_{1,i}}$;\\
{\small 7}\hspace{1.25cm} \textbf{If} $\nu = 2$ \textbf{then};\\
{\small 8}\hspace{1.7cm} $Q_{2,i} = x_{2,i} - x_{2,i}^{\text{(ref)}} - \kappa_{1,i}$;\\
{\small 9}\hspace{1.7cm} $\dot{\hat{\phi}}_{2,i} = -{k_{2,i}}{\sigma_{2,i}}\hat{\phi}_{2,i}+\frac{1}{2}\epsilon_{2,i}{k_{2,i}}|{Q_{2,i}}|^2$;\\
{\small 10}\hspace{1.6cm} ${\kappa_{2,i}} = - \frac{1}{2} (\delta_{2,i}+\epsilon_{2,i}\hat{\phi}_{2,i}){Q_{2,i}}$;\\
{\small 11}\hspace{1.15cm} \textbf{Else} \textbf{then}\\
{\small 12}\hspace{1.6cm} $Q_{\nu,i} = x_{\nu,i} - x_{\nu,i}^{\text{(ref)}}$;\\
{\small 13}\hspace{1.6cm} $\dot{\hat{\phi}}_{\nu,i} = -{k_{\nu,i}}{\sigma_{\nu,i}}\hat{\phi}_{\nu,i}+\frac{1}{2}\epsilon_{\nu,i}{k_{\nu,i}}|{Q_{\nu,i}}|^2$;\\
{\small 14}\hspace{1.6cm} ${\kappa_{\nu,i}} = - \frac{1}{2} (\delta_{\nu,i}+\epsilon_{\nu,i}\hat{\phi}_{\nu,i}){Q_{\nu,i}}$;\\
{\small 15}\hspace{1.2cm} \textbf{end}\\
{\small 16}\hspace{0.65cm} \textbf{end}\\
{\small 17}\hspace{0.1cm} \textbf{end}\\
\\
\hline
\end{tabular}
\end{table}
\subsection{Stability Analysis}
\label{sec:stability}
By considering the error of the adaptive law provided in \eqref{equation: 45}, as $\tilde{\phi}_{\nu,i} = \hat{\phi}_{\nu,i} - {\phi}^*_{\nu,i}$ in which ${\phi}^*_{\nu,i}$ is a positive constant depending on external disturbance and uncertainty bounds, we have:
\begin{equation}
\begin{aligned}
\label{equation: 450}
\dot{\tilde{\phi}}_{\nu,i} =& -{k_{\nu,i}}{\sigma_{\nu,i}}\hat{\phi}_{\nu,i}+\frac{1}{2}\epsilon_{\nu,i}{k_{\nu,i}}|{Q_{\nu,i}}|^2-{k_{\nu,i}}{\sigma_{\nu,i}}{\phi}^*_{\nu,i}
\end{aligned}
\end{equation}

For each EMLA, we can have four subsystem-based different Lyapunov functions, as follows:
\begin{equation}
\begin{aligned}
\label{56}
& V_i=\frac{1}{2} \sum_{\nu=1}^4\left(Q_{\nu, i}^2+k_{\nu, i}^{-1} \tilde{\phi}_{\nu, i}^2\right)
\end{aligned}
\end{equation}

Furthermore, we can extend \eqref{56} into the whole robot with $n_a$ EMLAs, as $V=\sum_{i=1}^{n_a} V_i$, meaning that:  
\begin{equation}
\begin{aligned}
\label{55}
V=&\frac{1}{2} \sum_{i=1}^{n_a} \sum_{\nu=1}^4 {Q^2_{\nu,i}}+{k^{-1}_{\nu,i}}\tilde{\phi}^2_{\nu,i}
\end{aligned}
\end{equation}
After the derivative of \eqref{56} and \eqref{55}, and considering equation \eqref{equation: 450}, we obtain \cite{heydarishahna2023robust}:
 \begin{equation}
\begin{aligned}
\label{54}
\dot{V} \leq-\zeta V+\frac{1}{4} \sum_{i=1}^{n_a} \lambda_i^{-1} \Pi_i^2+\bar{\zeta}
\end{aligned}
\end{equation}
where $\lambda_i$ is a positive constant, $\bar{\zeta}$ is a positive constant depending on the bound of disturbances, $\Pi_i$ is a strictly positive function depending on the bound of uncertainties, and:
\begin{equation}
\begin{aligned}
\label{5400}
\zeta = \min _{i \in\{1,2, \ldots, n_a\}} \begin{cases}
 \min (\delta_{1,i}\hspace{0.1cm}, \hspace{0.1cm}...\hspace{0.1cm},\hspace{0.1cm} \delta_{4,i})\\
 \min (k_{1,i}  \hspace{0.1cm} \sigma_{1,i} \hspace{0.1cm}, \hspace{0.1cm}...\hspace{0.1cm},\hspace{0.1cm}k_{4,i}  \hspace{0.1cm} \sigma_{4,i})
\end{cases}
\end{aligned}
\end{equation}
For further information; see \cite{heydari2024robust}. Thus, by employing the proposed control, the whole system is uniformly exponentially stable.

\subsection{Case Study: 3-DoF EMLA-Driven Manipulator}
\label{case_study_control}
\begin{figure}[h]
\centering
\includegraphics[trim={0.0cm 0.0cm 0.0cm 0.0cm},clip,width=5cm]{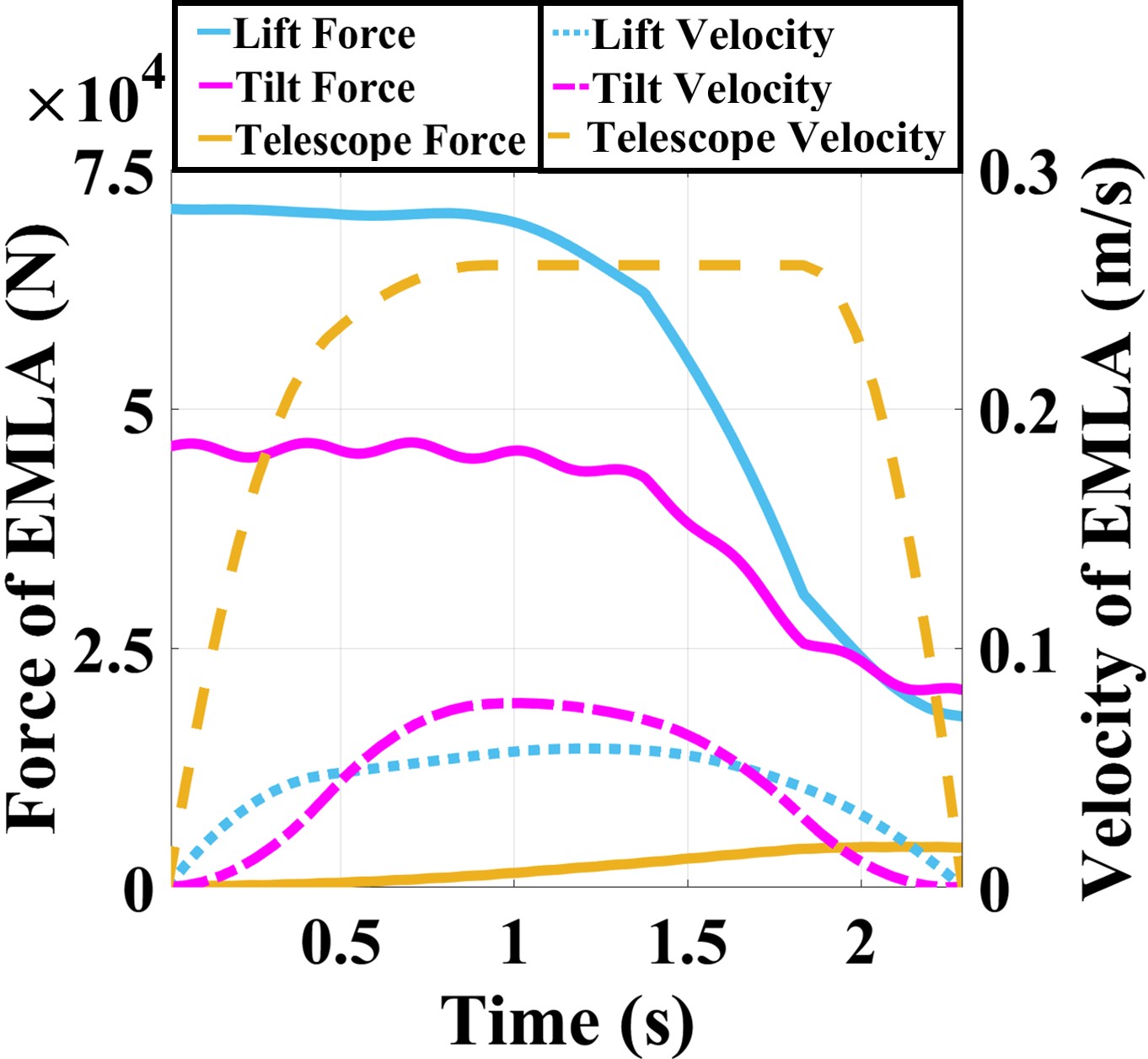}
\caption{\textcolor{black}{Controlled linear force and velocity of the EMLAs}}
\label{controlled_signals_mech}
\end{figure}

\begin{figure}[h]
\centering
\includegraphics[trim={0.0cm 0.0cm 0.0cm 0.0cm},clip,width=5cm]{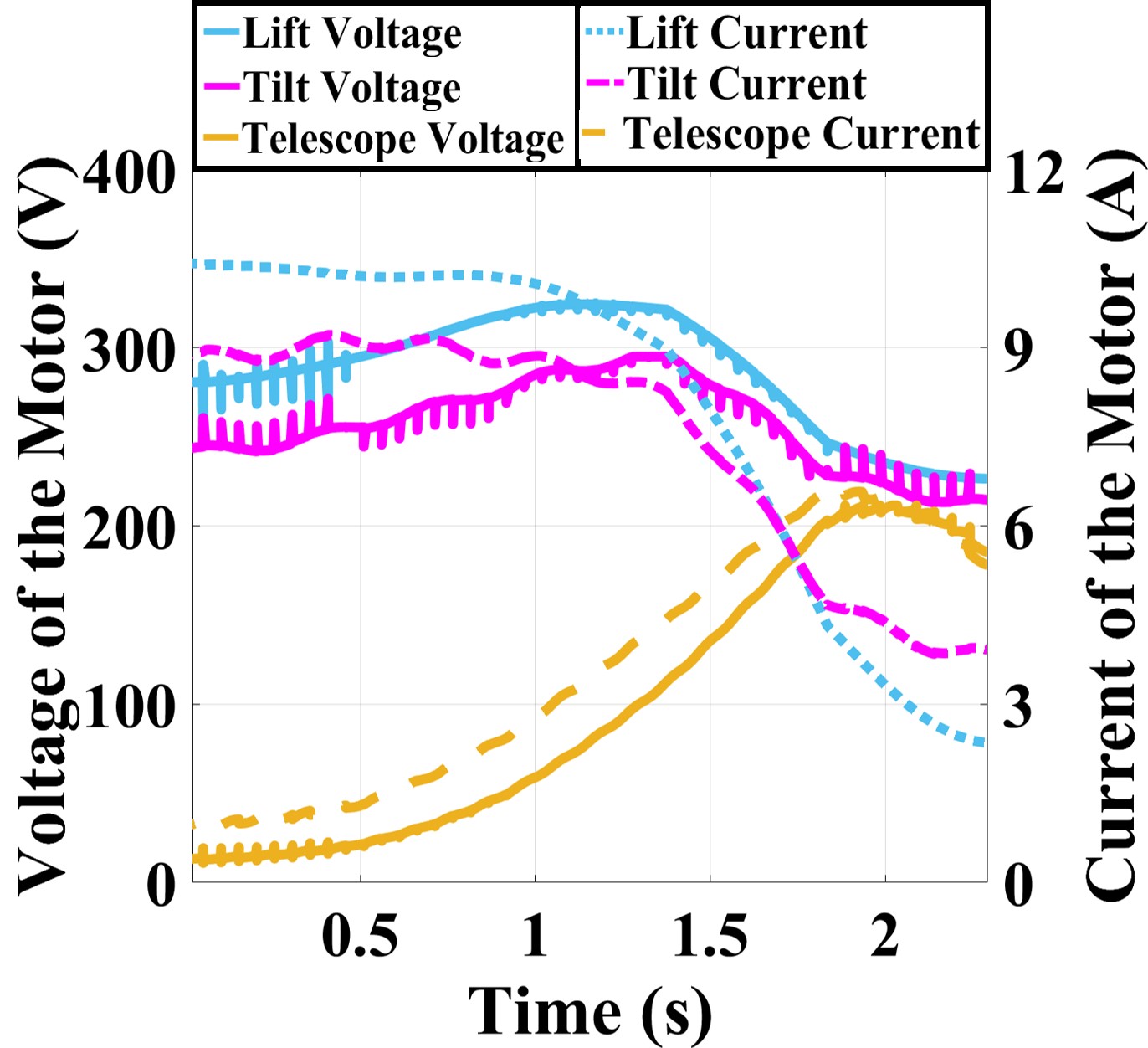}
\caption{\textcolor{black}{Controlled voltages and currents of the PMSMs}}
\label{controlled_signals_elec}
\end{figure}

\begin{figure}[h]
\centering
\includegraphics[trim={0.0cm 0.0cm 0.0cm 0.0cm},clip,width=5cm]{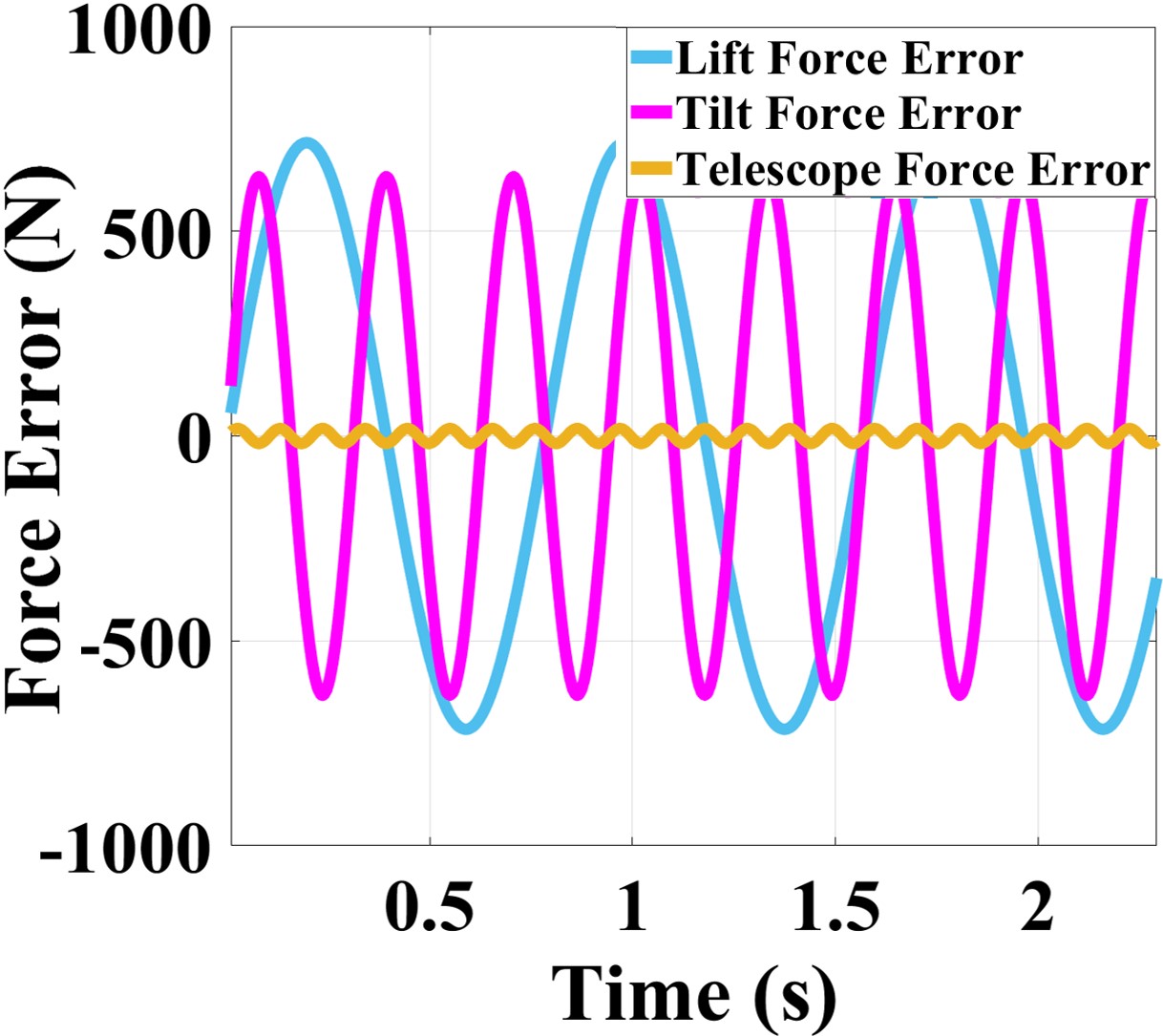}
\caption{\textcolor{black}{Force error of the EMLAs}}
\label{error_force}
\end{figure}

\begin{figure}[h]
\centering
\includegraphics[trim={0.0cm 0.0cm 0.0cm 0.0cm},clip,width=5cm]{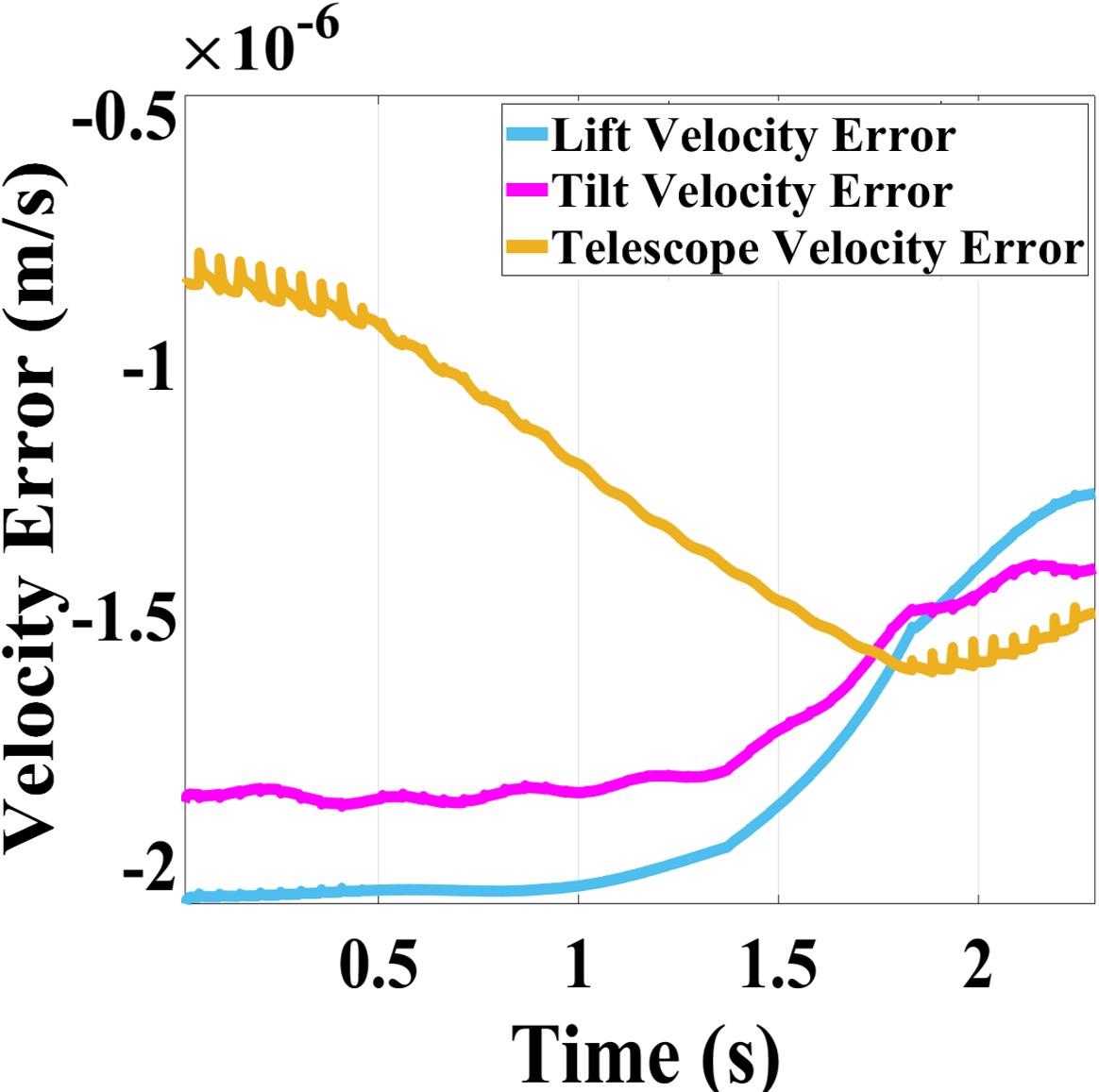}
\caption{\textcolor{black}{Velocity error of the EMLAs}}
\label{error_velocity}
\end{figure}

The fulfillment of the control objectives outlined in this paper involves the concurrent motion control of EMLAs incorporated in manipulator joints. In this case, we consider $n_a=3$, and three PMSMs will receive information regarding sufficient torque reference and subsequently control $i_q$ and $i_d$ by inducing adequate voltages to the stator windings to track the optimal motion reference. The control gains for all three EMLAs were selected as $k_{\nu,i}=7$, $\sigma_{\nu,i}=9$, $\delta_{\nu,i}=75000$, and $\epsilon_{\nu,i}=9$. As shown in Fig. \ref{controlled_signals_mech}, the controlled voltages and currents of the PMSMs (depicted in Fig. \ref{controlled_signals_elec}) provide adequate accuracy for the forces and velocities to closely track the reference values across all three EMLAs. The values of the controlled forces and velocities exhibit slight error, as shown in Fig. \ref{error_force} and Fig. \ref{error_velocity}, yet they remain effective in meeting the control objectives and stability in the presence of time- and state-variant disturbances.
\section{Conclusion}
This paper presented a bilevel multi-objective optimization framework aimed at enhancing the performance and energy efficiency of EMLA-driven heavy-duty manipulators in all-electric HDMMs. Through a comprehensive study that integrates the dynamics of implemented EMLAs with the heavy-duty manipulator one, the proposed framework effectively addresses the dual objectives of maximizing actuator efficiency and optimizing manipulator performance. To do this, this work developed a leader--follower optimization approach that harmonizes the operational dynamics between EMLAs and manipulators. By structuring the optimization problem into two interconnected levels, where the leader level focuses on maximizing the EMLA's efficiency and the follower level generates optimal trajectories for the manipulator, the framework ensures a balanced and synergetic operation of the entire system. To validate the proposed framework, a 3-DoF manipulator was considered. The results demonstrated substantial efficiency gains while maintaining high-performance operation. The actuation system achieved a total efficiency of 70.3$\%$, highlighting the effectiveness of the optimization approach in reducing energy consumption and extending the operational time of the HDMM. Moreover, the paper utilized a robust, adaptive, subsystem-based control strategy designed to ensure precise tracking of generated optimal trajectory, even in the presence of uncertainties and external disturbances. This control strategy not only improves the reliability and stability of the system but also simplifies its integration into existing industrial applications due to its modularity, thereby facilitating faster deployment and reducing maintenance costs. In conclusion, this study provides a practical solution for industries transitioning towards sustainable, electrified HDMM systems. By leveraging the proposed optimization framework and control strategy, HDMM manufacturers can achieve substantial gains in energy efficiency and minimize downtime. 

\section*{Appendix}
\label{ap:dynamic}
$\mathbf{M}_{\mathbf{A}} \in \mathbb{R}^{6 \times 6}$ represents the mass matrix, $\mathbf{C}_{\mathbf{A}} \! \left({ }^{\mathbf{A}} \omega\right) \in \mathbb{R}^{6 \times 6}$ denotes the matrix of Coriolis and centrifugal terms, and $\mathbf{G}_{\mathbf{A}} \in \mathbb{R}^{6}$ corresponds to the gravity terms and obtained as \eqref{eq:mass_matrix}-\eqref{gravity}.
\begin{equation}
\mathbf{M}_{\mathbf{A}}=\left[\begin{array}{ll}
m_{\mathbf{A}} \mathbf{I}_3 & -m_{\mathbf{A}}\left({ }^{\mathbf{A}} \mathbf{r}_{\mathbf{A B}} \times\right) \\
m_{\mathbf{A}}\left({ }^{\mathbf{A}} \mathbf{r}_{\mathbf{A B}} \times\right) & \mathbf{I}_{\mathbf{A}}-m_{\mathbf{A}}\left({ }^{\mathbf{A}} \mathbf{r}_{\mathbf{A B}} \times\right)^2
\end{array}\right]
\label{eq:mass_matrix}
\end{equation}
\begin{equation}
\begin{aligned}
& \mathbf{C}_{\mathbf{A}}\left({ }^{\mathbf{A}} \omega\right)=\left[\begin{array}{ll}
m_{\mathbf{A}}\left({ }^{\mathbf{A}} \boldsymbol{\omega} \times\right) & \\
m_{\mathbf{A}}\left({ }^{\mathbf{A}} \mathbf{r}_{\mathbf{A B}} \times\right)\left({ }^{\mathbf{A}} \boldsymbol{\omega} \times\right) & \left({ }^{\mathbf{A}} \boldsymbol{\omega} \times\right) \mathbf{I}_{\mathbf{A}} 
\end{array}\right. \\
& \begin{array}{l}
-m_{\mathbf{A}}\left({ }^{\mathbf{A}} \boldsymbol{\omega} \times\right)\left({ }^{\mathbf{A}} \mathbf{r}_{\mathbf{A B}} \times\right) \\
\left.+\mathbf{I}_{\mathbf{A}}\left({ }^{\mathbf{A}} \boldsymbol{\omega} \times\right)-m_{\mathbf{A}}\left({ }^{\mathbf{A}} \mathbf{r}_{\mathbf{A B}} \times\right)\left({ }^{\mathbf{A}} \boldsymbol{\omega} \times\right)\left({ }^{\mathbf{A}} \mathbf{r}_{\mathbf{A B}} \times\right)\right]
\end{array} \\
&
\end{aligned}
\label{eq:coriolis}
\end{equation}
\begin{equation}
\mathbf{G}_{\mathbf{A}}=\left[\begin{array}{l}
m_{\mathbf{A}}{ }^{\mathbf{A}} \mathbf{R}_{\mathbf{I}} \mathbf{g} \\
m_{\mathbf{A}}\left({ }^{\mathbf{A}} \mathbf{r}_{\mathbf{A B}} \times\right)^{\mathbf{A}} \mathbf{R}_{\mathbf{I}} \mathbf{g}
\end{array}\right]
\label{gravity}
\end{equation}
where $\mathbf{I}_3 \in \mathbb{R}^{3 \times 3}$ is the identity matrix, $m_{\mathbf{A}} \in \mathbb{R}$ represents the mass of the rigid body, $\mathbf{g}=[0,0,9.81]^T \in \mathbb{R}^3$ denotes the gravitational vector, and $\mathbf{I}_{\mathbf{A}}={ }^{\mathbf{A}} \mathbf{R}_{\mathbf{I}} \mathbf{I}_o(t) { }^{\mathbf{I}} \mathbf{R}_{\mathbf{A}}$ remains time-invariant.




\begin{IEEEbiography}[{\includegraphics[width=1in,height=1.25in,clip,keepaspectratio]{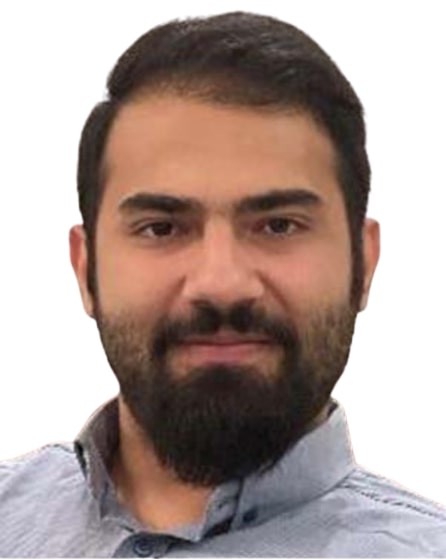}}]{Mohammad Bahari} earned a B.Sc. in electrical engineering power from Semnan University, Semnan, Iran, in 2015, followed by the completion of his M.Sc.  in electrical engineering power electronics and electric machines from Sharif University of Technology, Tehran, Iran, in 2019. Presently, he is engaged as a doctoral researcher at Tampere University, Tampere, Finland, focusing on design and control of an all-electric robotic e-boom. His research interests include multidisciplinary design optimization of electromechanical actuators, battery electric vehicles, and high-precision electromagnetic sensors.  
\end{IEEEbiography}

\begin{IEEEbiography}[{\includegraphics[width=1in,height=1.25in,clip,keepaspectratio]{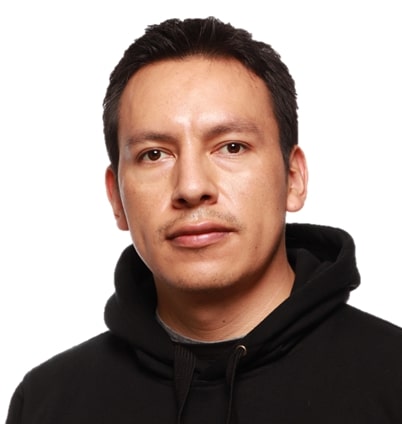}}]{Álvaro Paz Anaya} obtained his bachelor's degree in Electronics Engineering in 2013 from Instituto Tecnológico de San Juan del Río, Querétaro, Mexico. He received the M.Sc. and PhD degrees in Robotics and Advanced Manufacturing from Centro de Investigación y de Estudios Avanzados del IPN (CINVESTAV), Saltillo, Mexico in 2017 and 2022. He is currently a post doctoral researcher at Tampere University, Tampere, Finland. His research interests include robot trajectory optimization, humanoid whole body motion generation, multibody dynamic algorithms and mobile robotics.
\end{IEEEbiography}

\begin{IEEEbiography}[{\includegraphics[width=1in,height=1.25in,clip,keepaspectratio]{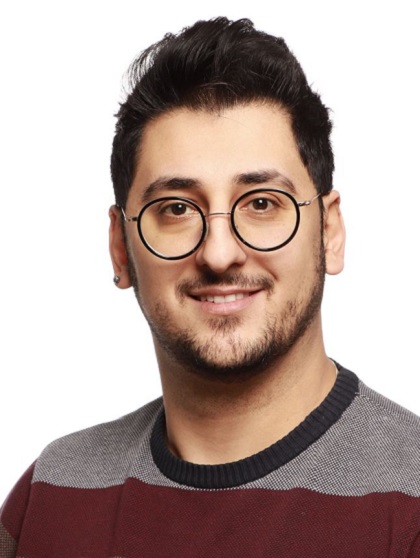}}]{Mehdi Heydari Shahna} earned a B.Sc. in electrical engineering from Razi University, Kermanshah, Iran, in 2015 and an M.Sc. in control engineering at Shahid Beheshti University, Tehran, Iran, in 2018. Since December 2022, he has been pursuing his doctoral degree in automation technology and mechanical engineering at Tampere University, Tampere, Finland. His research interests encompass robust control, nonlinear control of robotic systems, control of heavy-duty manipulators, fault-tolerant algorithms, and stability.
\end{IEEEbiography}

\begin{IEEEbiography}[{\includegraphics[width=1in,height=1.25in,clip,keepaspectratio]{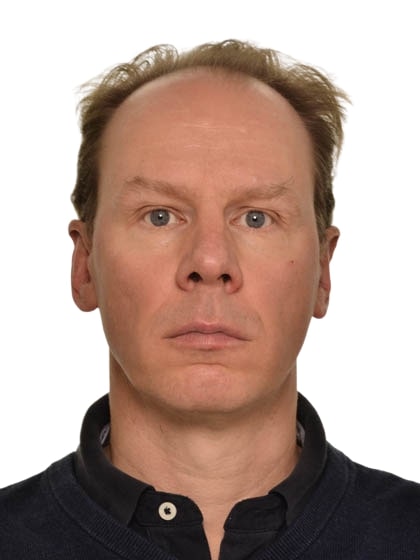}}]{Jouni Mattila}
received an M.Sc. and Ph.D. in automation engineering from Tampere University of Technology, Tampere, Finland, in 1995 and 2000, respectively. He is currently a professor of machine automation with the Unit of Automation Technology and Mechanical Engineering at Tampere University. His research interests include machine automation, nonlinear-model-based control of robotic manipulators, and energy-efficient control of heavy-duty mobile manipulators.
\end{IEEEbiography}

\vfill

\end{document}